\documentclass{osa-article}
\usepackage{gensymb}

\journal{osajournal}

\articletype{Research Article}

\begin{document}

\title{High-speed multiview imaging approaching 4pi steradians using conic section mirrors: theoretical and practical considerations}

\author{Kevin C. Zhou,\authormark{1,*} Al-Hafeez Dhalla,\authormark{1} Ryan P. McNabb,\authormark{2} Ruobing Qian,\authormark{1} Sina Farsiu,\authormark{1,2} and Joseph A. Izatt\authormark{1,2}}

\address{\authormark{1}Department of Biomedical Engineering, Duke University, Durham, NC 27708\\
\authormark{2}Department of Ophthalmology, Duke University Medical Center, Durham, NC 27708\\
}

\email{\authormark{*}kevin.zhou@duke.edu}



\begin{abstract}
Illuminating or imaging samples from a broad angular range is essential in a wide variety of computational 3D imaging and resolution-enhancement techniques, such as optical projection tomography (OPT), optical diffraction tomography (ODT), synthetic aperture microscopy, Fourier ptychographic microscopy (FPM), structured illumination microscopy (SIM), photogrammetry, and optical coherence refraction tomography (OCRT). The wider the angular coverage, the better the resolution enhancement or 3D resolving capabilities. However, achieving such angular ranges is a practical challenge, especially when approaching $\pm$90$^{\circ}$ or beyond. Often, researchers resort to expensive, proprietary high numerical aperture (NA) objectives, or to rotating the sample or source-detector pair, which sacrifices temporal resolution or perturbs the sample. Here, we propose several new strategies for multi-angle imaging approaching 4pi steradians using concave parabolic or ellipsoidal mirrors and fast, low rotational inertia scanners, such as galvanometers. We derive theoretically and empirically relations between a variety of system parameters (e.g., NA, wavelength, focal length, telecentricity) and achievable fields of view (FOVs) and importantly show that intrinsic tilt aberrations do \textit{not} restrict FOV for many multi-view imaging applications, contrary to conventional wisdom. Finally, we present strategies for avoiding spherical aberrations at obliquely illuminated flat boundaries. Our simple designs allow for high-speed multi-angle imaging for microscopic, mesoscopic, and macroscopic applications.
\end{abstract}

\section{Introduction}
Angular diversity is important in a wide variety of imaging techniques as a means for enhancing  not only lateral resolution, but also axial resolution to enable high-resolution 3D imaging. Perhaps the most familiar examples are standard wide-field or point-scanning microscopy, whose 3D spatial resolutions improve when using objectives with increasing numerical aperture (NA). Whether detecting coherent scattering or incoherent fluorescence, higher NA objectives provide access to higher illumination or collection angles, which contain higher-spatial-frequency information. In these cases, the information from all angles is present simultaneously. A variation of this is imaging based on speckled illumination, whose theoretical resolution is proportional to speckle grain size, which in turn depends on angular diversity. Other techniques computationally combine images sequentially acquired from multiple illumination angles to create a synthetic aperture, thereby reconstructing a 3D or resolution-enhanced image. Examples include optical diffraction tomography (ODT) \cite{wolf1969three, lauer2002new, chowdhury2019high, zhou2020diffraction}, Fourier ptychographic microscopy (FPM) \cite{zheng2013wide, konda2020fourier, zheng2021concept}, and structured illumination microscopy (SIM) \cite{gustafsson2000surpassing, chowdhury2017refractive}, including with speckled illumination \cite{mudry2012structured, dong2014high, yeh2019speckle}. Both the illumination and collection angles may also be varied, as in optical projection tomography (OPT) \cite{sharpe2002optical}, optical coherence projection tomography (OCPT) \cite{van2020deep}, and optical coherence refraction tomography (OCRT)  \cite{zhou2019optical, zhou2020spectroscopic}, which can be achieved through rotation of the sample or source-detector pair. While OPT and OCPT are transmissive modalities, OCRT is reflective and thus illumination steering automatically steers the detector. Finally, yet another class of related techniques, including photogrammetry, structure-from-motion (SfM), and multi-view stereo (MVS) \cite{ullman1979interpretation, wu2013towards,furukawa2015multi,schonberger2016structure,zhou2021mesoscopic}, reconstructs a 3D surface using a sequence of non-telecentric images (e.g., from standard consumer-grade cameras) acquired from multiple angles.

In all of these techniques, the wider the angular coverage, whether across a physical or synthetic aperture, the better the theoretical resolution. At the same time, however, the wider the angular coverage, the more difficult it is to achieve, especially when approaching $\pm$90$^{\circ}$. While high-NA objectives are commonly used, they are in practice limited to approximately $\pm$70$^{\circ}$ (i.e., NA=0.95 in air objectives) over small fields of view (FOVs) and only allow multi-angle illumination of samples from one side. As such, techniques like OPT and photogrammetry typically employ mechanical rotation of the sample or detector, which sacrifices temporal resolution or require contrived sample preparation.

\begin{figure}
    \centering
    \includegraphics[width=.6\linewidth]{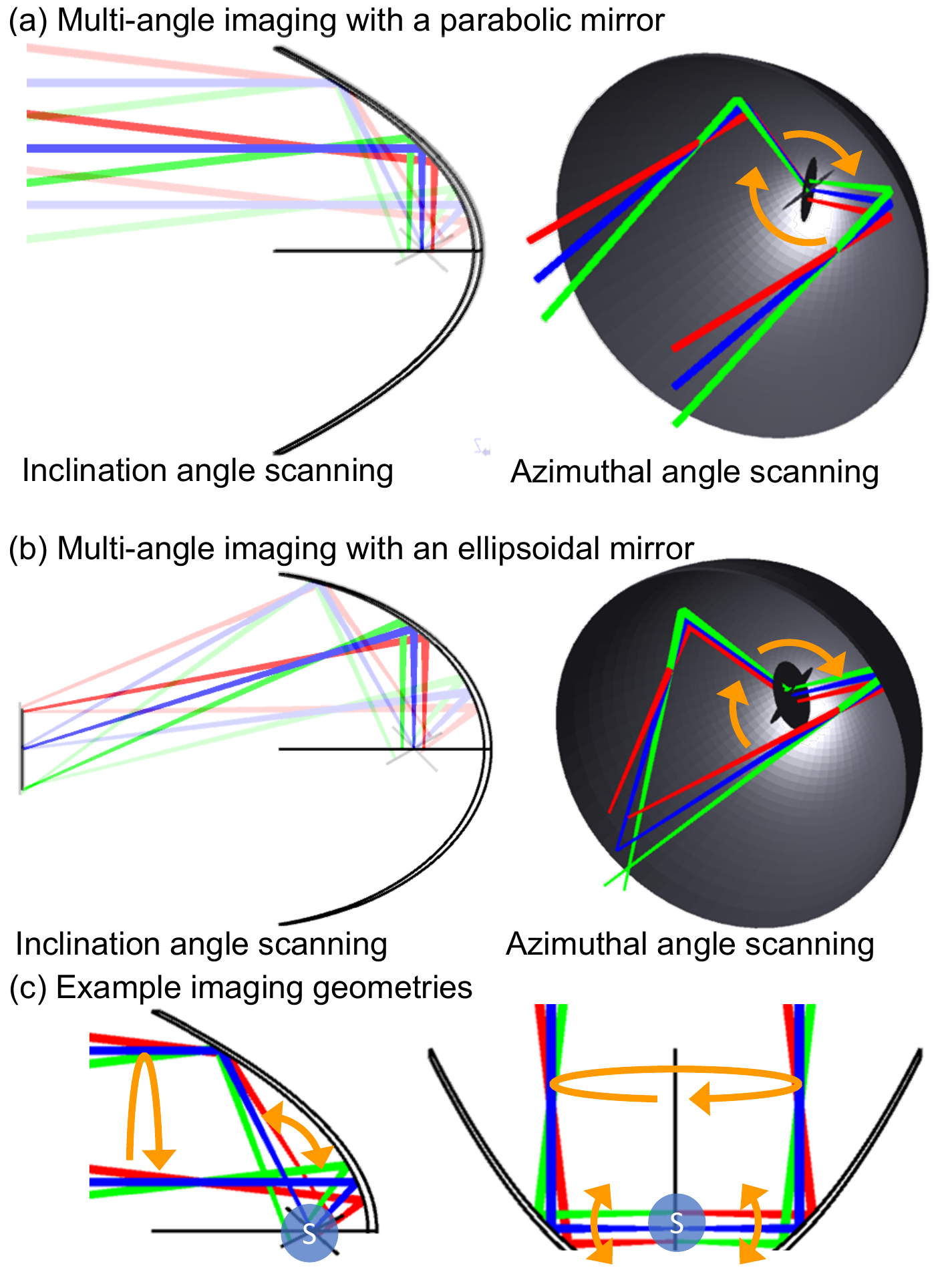}
    \caption{Parabolic (a) and ellipsoidal (b) mirrors with large, rotationally symmetric apertures (or half apertures) can enable imaging from wide angular ranges across one or two axes using different beam scanning mechanisms (S: sample). Example wide-angle imaging designs are shown in (c). The azimuthal angular range can theoretically reach 180$^\circ$ for a half aperture or 360$^\circ$ for a full aperture, while the inclination angular range depends on the aperture size. Only one of the two lateral scan dimension is shown for simplicity. Note that the image plane is tilted and orthogonal to the central chief ray.
    }
    \label{fig1}
\end{figure}

Here, we propose several different designs that can achieve multi-angle imaging approaching 4pi steradians, without the need for rotating the sample or source-detector pair. Our proposed designs achieve this by employing low rotational inertial scanning (e.g., of galvanometers) across the apertures of conic section mirrors, such as parabolic, ellipsoidal, or axicon mirrors. Conceptually equivalently, we can altogether avoid mechanical scanning by using a multi-camera array design \cite{lin2015camera,fan2019video,zhou2021snapshot}.  Parabolic and ellipsoidal mirrors have been previously incorporated in a wide variety of imaging systems, such as confocal microscopy \cite{lieb2001high, drechsler2001confocal}, multiphoton microscopy \cite{combs2007optimization},  total internal reflection fluorescence (TIRF) microscopy \cite{ruckstuhl2004attoliter, liu2017elliptical}, Raman imaging \cite{zhang2009parabolic}, photoacoustic imaging \cite{alshahrani2019all}, FPM \cite{lee2019reflective}, and the recently proposed random access microscopy \cite{ashraf2021random}. While the primary attractive feature of these conic section mirrors is their access to very wide angular ranges and their ability to focus ``perfectly'' at very high NAs, they exhibit well known tilt aberrations that restrict high-NA focusing to very small regions. As a result, many of these imaging techniques require physically translating the sample to form an image \cite{lieb2001high, drechsler2001confocal, ruckstuhl2004attoliter, zhang2009parabolic} or use conic section mirrors only for illumination \cite{alshahrani2019all, liu2017elliptical, lee2019reflective}. Few techniques use these mirrors for 2D image formation as an imaging objective, which requires using tilted rays that lead to aberrations. This usage is of primary interest and explored in detail in this paper. 

To this end, building upon previous works that study the effects of input illumination tilt on focus quality \cite{howard1979imaging, arguijo2003exact, april2011focusing, labate2016effects, zeng2020far}, we carefully characterize the achievable FOVs over which near-diffraction-limited performance is achievable as a function of various experimental parameters (key equations: Eqs. \ref{FOV_x} and \ref{FOV_x_quad}). These parameters include NA, mirror parameters (e.g., focal length, major and minor axes), sample-incident angle, wavelength, and chief ray geometry (i.e., whether a telecentric design is used). Crucially, we will argue that for 3D imaging applications, the well-known tilt aberrations do \textit{not} significantly limit the lateral FOV, any more than the depth of field limits the axial FOV. This work lays the theoretical foundation that generalizes the parabolic-mirror-based 3D imaging method for which we recently presented preliminary results \cite{zhou2021incoherent}.

\section{Wide-angle multi-view imaging with conic section mirrors} \label{sec2}
We first describe the general strategies for imaging over wide angular ranges, common to all conic section mirror shapes considered here, which achieve similar results slightly differently (Fig. \ref{fig1}). Specifically, our goal is to acquire 2D (or 3D) images from multiple views over a very wide angular range, whether with active illumination via point-scanning or with full-field detection with a 2D camera. Our explanations proceed based on a point-scanning system, with the understanding that similar rays govern a 2D-camera-based imaging system (Sec. \ref{camera}). Further, this concept is applicable to both reflection and transmissive imaging geometries (Sec. \ref{trans_refl}).

To obtain multi-view imaging over 360$^{\circ}$ about a single axis, we propose using a rotationally symmetric, concave mirror as a focusing element that reflects incident beams and weakly focuses them at a focal point, where the sample is placed (Fig. \ref{fig1}). Since the mirror is rotationally symmetric, varying the incidence position along a circle concentric to the mirror aperture allows varying the incidence angle to the sample to cover the full 360$^{\circ}$. An image is formed by independently scanning the focus laterally, the mechanism of which depends on the mirror properties.

This idea can be straightforwardly extended to two rotational axes by simply allowing the entry position across the mirror aperture to vary independently over two dimensions. The angular range of the second rotation axis is asymmetric and slightly more limited, depending on the aperture size relative to the focal length. In particular, as the entry position approaches the optic axis, the sample-incident angle approaches 90$^{\circ}$, which may be clipped if the incident beam is blocked by the sample before hitting the mirror. Similarly, as the entry position moves away from the optic axis, the incidence angle increases in the opposite direction, and achieves a maximum angle limited by the mirror aperture size or input scanning mechanism. Thus, if the sample is small and the aperture is very large, nearly 4pi steradians of coverage is theoretically possible. In practice, sample mounting constraints may restrict the range to, for instance, $\pm$90$^{\circ}$ or 2pi steradians, if we strictly only have access to one side of the sample (Fig. \ref{fig1}c). 

A number of system designs can achieve these aforementioned multi-angle imaging properties, but we focus on two concave conic section mirrors, parabolic and ellipsoidal mirrors, due to their commercial availability and conceptual simplicity. In the following sections, we discuss in detail the properties of these two mirrors and how they can achieve multi-view imaging over wide angular ranges.





\section{Parabolic mirror}\label{sec3}

\begin{figure}
    \centering
    \includegraphics[width=.8\linewidth]{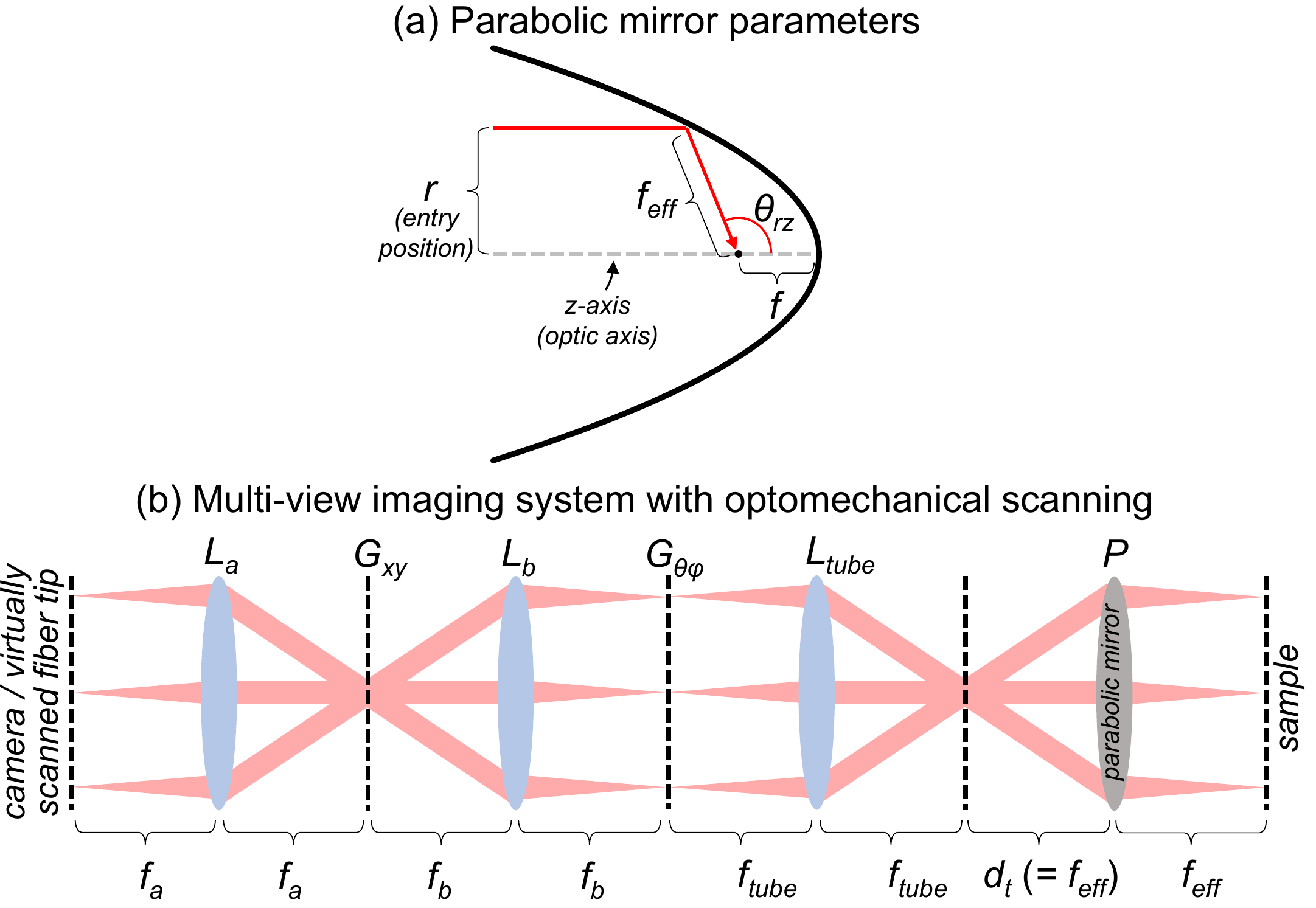}
    \caption{(a) Parabolic mirror parameters used in Sec. \ref{sec3}. (b) Generic imaging system design using a pair of anti-conjugate galvanometers to achieve spatio-angular beam scanning for multi-view imaging. Lenses a and b ($L_a$ and $L_b$) form a 4f relay, with focal lengths $f_a$ and $f_b$, that enforce anti-conjugate and conjugate galvanometers, $G_\mathit{xy}$ and $G_\mathit{\theta\phi}$, which scan lateral position at the sample and the sample-incident angle, respectively. Alternatively, if a camera is used, $G_\mathit{xy}$ is replaced with a stop that dictates the lateral resolution of the system.
    The tube lens $L_\mathit{tube}$ (focal length $f_\mathit{tube}$) joins the parabolic mirror $P$ in another 4f relay, the latter of which is governed by the effective focal length $f_\mathit{eff}$ (not $f$). The spacing between the tube lens and parabolic mirror ($d_t$) can be varied to adjust the telecentricity of the laterally scanned beam at the sample. When $d_t=f_\mathit{eff}$, it is telecentric.}
    \label{fig:parabolic}
\end{figure}

Parabolic (technically, paraboloidal) mirrors are often thought of as ``ideal'' focusing elements for collimated beams. Unlike the more common spherical optics, they can achieve aberration-free, diffraction-limited focusing of untilted (parallel to the optic axis) beams, regardless of the input beam diameter or position. A parabolic mirror surface $P$ can be parameterized by a single value, the focal length $f$, as
\begin{equation}
    P(r)=\frac{r^2}{4f},
\end{equation}
where $r$ is the radial entry position along the mirror aperture. By definition, $f$ is the distance from the mirror apex to the nominal focus, where all collimated, untilted incident rays converge and where the sample is placed. In particular, the sample-incident angle in the $rz$-plane, relative to the $z$-axis (the mirror optic axis), after reflection off of the mirror (Fig. \ref{fig:parabolic}a), is given by
\begin{equation}\label{para_theta}
    \theta_\mathit{rz}(r)=2\tan^{-1}\left(\frac{r}{2f}\right).
\end{equation}
Thus, the maximum inclination angular range is restricted by the mirror aperture size (relative to its focal length).
Note that the distance between the focus and where a ray intersects with the mirror surface also varies as a function of $r$, meaning that the lateral resolution varies with $r$. Thus, it is useful to define this distance as the effective focal length, 
\begin{equation} \label{f_eff}
    f_\mathit{eff}(r)=f+\frac{r^2}{4f}.
\end{equation}
The effective output NA in air is thus given by
\begin{equation}\label{NAeff}
    \mathit{NA}(r)=
    \sin\left(\frac{\theta_\mathit{rz}(r+w/2)-\theta_\mathit{rz}(r-w/2)}{2}\right)
    \approx\sin\left(\frac{w}{2f_\mathit{eff}(r)}\right)
    \approx\frac{w}{2f_\mathit{eff}(r)},
\end{equation}
where $w$ is the input beam diameter. In both approximations, we assume that $w\ll f$, first in the argument of the sine, and then in the overall expression. In this small-angle regime, the $\mathit{NA}$ of the parabolic mirror is a simple function of $f_\mathit{eff}$. For the single-axis 360$^{\circ}$ imaging configuration described in Sec. \ref{sec2}, the incidence position is varied azimuthally with a fixed $r$ and therefore fixed $f_\mathit{eff}$ and $\mathit{NA}$. In particular, when the special case of 
\begin{equation}\label{para_special}
    f_\mathit{eff}^{90^{\circ}}=2f
\end{equation}
is met, the sample-incidence angle $\theta_\mathit{rz}=90^{\circ}$, meaning the multi-angle illumination central chief rays sweep out a plane. However, as the second rotation axis is achieved by varying $r$, the resolution necessarily varies as well, assuming $w$ is not dynamically adjusted.

So far, we have only described focusing of incident beams parallel to the optic axis, such that the beams always converge at the mirror's nominal focus. To scan the focus laterally and form a 2D image, we scan the beam's incident angle to the mirror aperture. The pivot point of this beam angle scanning determines whether the lateral scanning is telecentric (parallel chief rays) or non-telecentric (fanned chief rays, as in the human visual systems and most consumer cameras). To understand this point, it's instructive to think of the parabolic mirror as part of a 4f imaging relay that includes a second focusing element, a tube lens with focal length $f_\mathit{tube}$, to assist in input beam angular scanning (Fig. \ref{fig:parabolic}b). The 4 focal lengths are thus $f_\mathit{eff}+f_\mathit{eff}+f_\mathit{tube}+f_\mathit{tube}$, meaning a spacing of 
\begin{equation}\label{parabolic_telecentric}
    d(r)=f_\mathit{eff}(r)+P(r)+f_\mathit{tube}
\end{equation}
between the parabolic mirror apex and the tube lens' principal plane. Note that telecentricity can only be achieved for one value of $r$, and therefore one sample-incident angle $\theta_\mathit{rz}$ and $f_\mathit{eff}$, without adjusting element spacing. A pair of beam-steering elements can be placed at image and Fourier planes (conjugate and anti-conjugate planes) of the sample to independently vary sample-incident position and angle (Fig. \ref{fig:parabolic}b), as described previously \cite{carrasco2015pupil}. For example, one might use 2D MEMS scanners or pairs of galvanometers, either imaged onto each other or slightly offset around the same focal plane. Another approach is to use a lens array, which achieves the same effect as scanning across the back aperture.

We can specify this simple imaging system's incident-angle-dependent theoretical magnification, as dictated by its 4f configuration, as
\begin{equation}
    M(r)\approx\frac{f_\mathit{tube}}{f_\mathit{eff}(r)},
\end{equation}
which is important for choosing scanning mirror sizes or sensor sizes in the case of non-point-scanning systems. However, the maximum achievable lateral FOV is much more challenging to analyze theoretically through basic optics principles, which is also generally the case for high-NA objectives. This is because the lateral FOV depends on aberrations, which are not present for the perfect focusing of untilted beams, but get progressively worse as the beam is tilted and scanned laterally. As we will show in a full analysis below (Sec. \ref{fov}), this fall-off in focusing quality depends on a variety of factors, including NA, the wavelength, telecentricity, and both $f$ and $f_\mathit{eff}$, independently.

\section{Ellipsoidal mirror}\label{sec4}

\begin{figure}
    \centering
    \includegraphics[width=.6\linewidth]{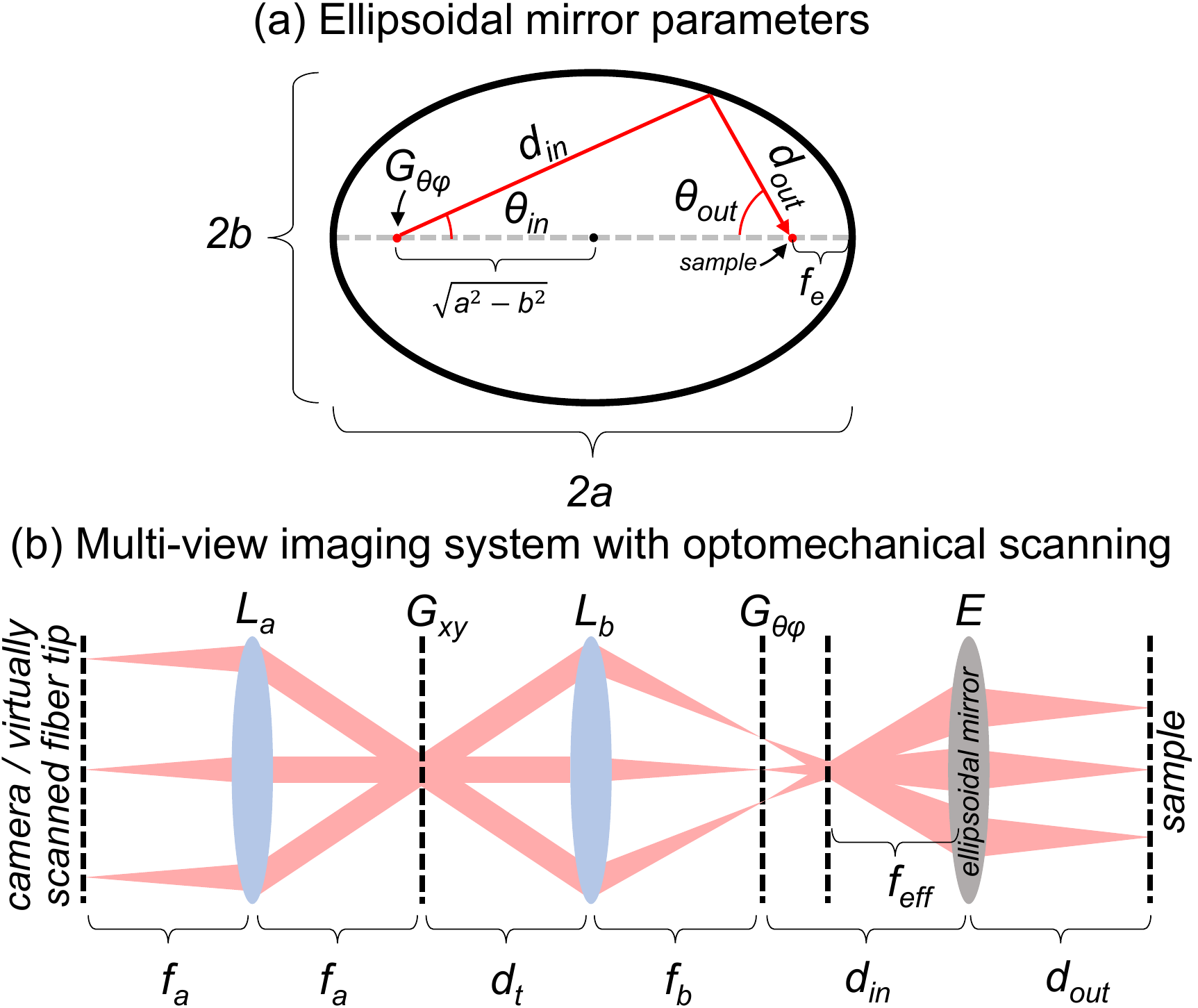}
    \caption{(a) Ellipsoidal mirror parameters used in Sec. \ref{sec4}. (b) Generic imaging system design using a pair of anti-conjugate galvanometers ($G_\mathit{xy}$ and $G_\mathit{\theta\phi}$) to achieve spatio-angular beam scanning for multi-view imaging. This design differs slightly from that of the parabolic mirror (Fig. \ref{fig:parabolic}), as the sample is not placed $f_\mathit{eff}$ away from ellipsoidal mirror ($E$). As a result, lenses a and b ($L_a$ and $L_b$) do not necessarily form a 4f system with their respective focal lengths, $f_a$ and $f_b$. By tuning the spacing between $L_a$ and $L_b$ via $d_t$, we can ensure telecentric lateral scanning at the sample (Eq. \ref{ellipse_telecentric}).}
    \label{fig:ellipsoidal}
\end{figure}

Ellipsoidal mirrors are also considered ``perfect'' focusing elements for diverging fields originating from the other focus of the mirror. Specifically, a symmetric ellipsoidal mirror $E$ can be parameterized by two values, the semi-major and semi-minor axes, $a$ and $b$ (Fig. \ref{fig:ellipsoidal}a), as
\begin{equation}
    E(r)=\pm a\sqrt{1-\frac{r^2}{b^2}},
\end{equation}
where $r$ is the coordinate along the minor axis, and the two halves (from the $\pm$) each envelope one of the two foci, which are located at
\begin{equation}
    f_\pm = \pm \sqrt{a^2-b^2},
\end{equation}
relative to the center of the ellipse. All rays originating from one focus intersect with the other focus, where the sample of interest is placed, after reflecting off of the mirror, with a fixed propagation distance of $2a$. If a ray originating from one focus is launched at an angle $\theta_\mathit{rz}^\mathit{in}$, defined relative to the $z$-axis, the distance the ray travels to the mirror surface is given by
\begin{equation}
    d_\mathit{in}(\theta_\mathit{rz}^\mathit{in})=\frac{a(1-e^2)}{1-e\cos(\theta_\mathit{rz}^\mathit{in})},
\end{equation}
where $e=f_+/a$ is the eccentricity of the ellipsoid. Thus, it is straightforward to calculate the distance from the mirror to the other focus as
\begin{equation}
    d_\mathit{out}(\theta_\mathit{rz}^\mathit{in})=2a-d_\mathit{in}(\theta_\mathit{rz}^\mathit{in}).
\end{equation}

Thus, an ellisoidal mirror can be thought of as a non-telecentric, finite-conjugate imaging system, with an effective focal length governed by the lens equation,
\begin{equation}\label{thinlens}
    \frac{1}{f_\mathit{eff}(\theta_\mathit{rz}^\mathit{in})}=\frac{1}{d_\mathit{in}(\theta_\mathit{rz}^\mathit{in})}+\frac{1}{d_\mathit{out}(\theta_\mathit{rz}^\mathit{in})}
\end{equation}
Further, the output angle $\theta_\mathit{rz}^\mathit{out}$ can be calculated using the law of sines as
\begin{equation} \label{theta}
    \theta_\mathit{rz}^\mathit{out}(\theta_\mathit{rz}^\mathit{in}) = \begin{cases} 
          \pi - \sin^{-1}\left(d_\mathit{in}/d_\mathit{out}\sin\left(\theta_\mathit{rz}^\mathit{in}\right)\right) & d_\mathit{out} \leq \frac{b^2}{a} \\
          \sin^{-1}\left(d_\mathit{in}/d_\mathit{out}\sin\left(\theta_\mathit{rz}^\mathit{in}\right)\right) & d_\mathit{out} > \frac{b^2}{a}
       \end{cases},
\end{equation}
where the switching condition of $d_\mathit{out}=b^2/a$ corresponds to a $\theta_\mathit{rz}^\mathit{out}=90^{\circ}$ incidence to the sample at the second focus (under this condition, $d_\mathit{out}$ is known as the semilatus rectum). This occurs when 
\begin{equation}
    \theta_\mathit{rz}^\mathit{in, 90^{\circ}}=\sin^{-1}\left(\frac{b^2}{2a^2-b^2}\right)
\end{equation}

For a given mirror, the maximum inclination angular range is theoretically unbounded, but in practice restricted by the angular scan range of the galvanometers at the other focus. By Eq. \ref{thinlens}, we have 
\begin{equation}\label{feff_ellipsoidal}
    f_\mathit{eff}^{90^{\circ}}=\frac{b^2(2a^2-b^2)}{2a^3}.
\end{equation}
In other words, this configuration corresponds to the 360$^{\circ}$ single-rotation-axis imaging configuration analogous to Eq. \ref{para_special} for parabolic mirrors.


To find how input $\mathit{NA}_\mathit{in}=\sin(\alpha)$ at the first focus maps to the output $\mathit{NA}$ at the second focus, we can use Eq. \ref{theta} to propagate the marginal rays, yielding
\begin{equation}\label{NAout}
    \mathit{NA}(\theta_\mathit{rz}^\mathit{in})=
    \sin\left(
    \frac{
    |\theta_\mathit{rz}^\mathit{out}(\theta_\mathit{rz}^\mathit{in}-\alpha)-
    \theta_\mathit{rz}^\mathit{out}(\theta_\mathit{rz}^\mathit{in}+\alpha)|}
    {2}\right)
    \approx\sin\left(\tan^{-1}\left(\frac{d_\mathit{in}}{d_\mathit{out}}\tan(\alpha)\right)\right)
    \approx\frac{d_\mathit{in}}{d_\mathit{out}}\alpha,
\end{equation}
where the first approximation assumes that the left and right marginal rays travel the same distance to the mirror (and the same distance from the mirror to the second focus), and the second approximation assumes a small $\alpha$.

Thus, just like parabolic mirrors, ellipsoidal mirrors can also achieve multi-view imaging over 360$^{\circ}$ single-axis or over two axes with the same physical constraints. One practical advantage of ellipsoidal mirrors is that they behave like finite-conjugate imaging systems and thus do not require a tube lens, while parabolic mirrors behave more like 4f imaging systems and thus require an extra tube lens. Nevertheless, it is possible to achieve object-side telecentricity through a judicious choice of lenses and their spacings (Fig. \ref{fig:ellipsoidal}b). In particular, we can use two lenses (lenses a and b) in an almost-4f configuration, varying the third f ($d_t$) to achieve two intermediate focal planes: one where the marginal rays converge and the other where the central chief rays converge. The former corresponds to the other focal point of the ellipsoidal mirror (the one without the sample), where the sample-incident-angle-scanning galvanometers ($G_\mathit{\theta\phi}$) are placed. The latter focal plane is $f_\mathit{eff}$ (Eq. \ref{thinlens}) away from the ellipsoidal mirror surface for object-side telecentric scanning. The $d_t$ that achieves this by satisfying all the imaging conditions can be shown to be
\begin{equation}\label{ellipse_telecentric}
    d_t=\frac{f_b(d_\mathit{in}^2 + d_\mathit{in}f_b+d_\mathit{out}f_b)}{d_\mathit{in}^2}.
\end{equation}

Finally, we can specify the approximate magnification onto the galvo placed at the other ellipsoidal focus, based on the thin-lens model:
\begin{equation}
    M(\theta_\mathit{rz}^\mathit{in})\approx
    \frac{ d_\mathit{in}(\theta_\mathit{rz}^\mathit{in})}{ d_\mathit{out}(\theta_\mathit{rz}^\mathit{in})}.
\end{equation}
As with the parabolic mirror case, we will defer discussion of FOV to Sec. \ref{fov}.

\section{Field of view and space-bandwidth product} \label{fov}
The FOV can be defined as the spatial lateral (or axial) range over which aberrations are ``acceptably'' small. We adopt the criterion where the Strehl ratio is greater than 0.8 (1 = aberrationless), though in practice lower values may still be usable at the cost of signal-to-noise ratio (SNR) and resolution. While FOV cutoffs are in general arbitrary (just like for resolution), their dependence on other system parameters remain well defined. Perhaps a familiar example of this is arbitrarily defining the axial FOV by the depth of field of a camera or depth of focus of a point-scanned Gaussian beam to be when the defocus, the simplest form of aberration, causes the spot size to expand by a factor of $\sqrt{2}$ (i.e., the Rayleigh range or confocal parameter). Regardless of the definition however, the axial FOV for Gaussian optics scales quadratically with the lateral resolution or output NA:
\begin{equation}\label{FOV_z}
    \mathit{FOV}_z\propto \frac{\delta_\mathit{x}^2}{\lambda}
    \propto
    \frac{\lambda}{\mathit{NA}^2}
    ,
\end{equation}
where $\lambda$ is the wavelength. While axial FOV is restricted by defocus, the lateral FOVs for conic section mirrors are restricted by higher-order aberrations, notably astigmatism and coma from ``misaligning'' the mirror from its perfect, untilted focusing conditions. Here, as described in the preceding sections, we are essentially using misalignment to perform lateral scanning. Extending previous studies on characterizing the effects of such aberrations in misaligned parabolic mirrors on focus quality \cite{howard1979imaging, arguijo2003exact, april2011focusing, labate2016effects, zeng2020far}, here we provide a thorough investigation of the dependence of the lateral FOV, $\mathit{FOV}_\mathit{x}$ (or $\mathit{FOV}_\mathit{y}$, which is approximately the same), on a variety of relevant system design parameters: lateral resolution ($\delta_\mathit{x}$ or $\delta_\mathit{y}$) or $\mathit{NA}$, wavelength ($\lambda$), mirror size (via $f$, or $a$ and $b$), entry position or sample incidence angle, and telecentric vs. non-telecentric designs. Unless otherwise noted, our results for $\mathit{FOV}_\mathit{x}$ and $\delta_\mathit{x}$ also apply identically to $\mathit{FOV}_\mathit{y}$ and $\delta_\mathit{y}$, respectively. Note that ``lateral'' here is defined relatively along the plane orthogonal to the view direction (i.e., the central chief ray), which may be oblique relative to the mirror optic axis after reflecting off of the mirror surface.

\subsection{Approximate theoretical predictions of FOV}
We first derive a partial theoretical model that illustrates the scalings of lateral FOV with the various design parameters for parabolic mirrors using analytical ray tracing. Alongside the diffraction-limited spot size, $\delta_{x}$, we define $\delta_g$ as the geometric spot size by the same arbitrary criterion by which $\delta_{x}$ is defined. Note that $\delta_{x}$ changes depending on $f_\mathit{eff}$. Then, the aberrated spot size that factors in diffraction can be modeled as \cite{lohmann1989scaling}
\begin{equation} \label{spot_size}
    \delta_a=\sqrt{\delta_x^2+\delta_g(\mathbf{r}_f)^2},
\end{equation}
noting that the geometric spot size depends on lateral position along the focal plane, $\mathbf{r}_f=(x_f,y_f)$, which is 0 (i.e., diffraction-limited) at the nominal focus of the mirror, $\delta_g(0,0)=0$. For defining FOV, we are interested in the range over which $\delta_a\approx\delta_x$, which requires a better understanding of how the geometric spot size varies with lateral position. This relationship itself also depends on the theoretical diffraction-limited spot size $\delta_x$ (or equivalently the input beam diameter $w$) and the parabolic mirror aperture entry position $x_m$ relative to the optic axis (here, equivalent to $r$ used in earlier sections, so that $r=\sqrt{x_m^2+y_m^2}$, with $y_m=0$ for simplicity). We consider two limiting scenarios: 1) $x_m\gg0$ (off-axis) and 2) $x_m=0$ (on-axis).

When $x_m\gg0$, we derive in the supplemental document for parabolic mirrors that the geometric spot size increases linearly with distance from the nominal focus due to aberrations, with slope
\begin{equation}\label{beta}
    \beta \propto \frac{x_m w}{ff_\mathit{eff}}.
\end{equation}
We use this result with Eq. \ref{spot_size} to estimate the FOV scaling. Substituting this newly derived linear relation of $\delta_g(\mathbf{r}_f)\propto\beta r_f$ (where $r_f=|\mathbf{r}_f|$), we obtain
\begin{equation}\label{taylor1}
    \delta_a(r_f)=\sqrt{\delta_x^2+\beta^2 r_f^2}\approx
    \delta_x+\frac{\beta^2 r_f^2}{2\delta_x},
\end{equation}
where we have taken the second-order Taylor expansion at $r_f=0$. Thus, near the mirror's nominal focus, the spot size is approximately diffraction-limited, and scales quadratically with lateral position. We define the lateral FOV, $\mathit{FOV}_r=\mathit{FOV}_x=\mathit{FOV}_y$ as the lateral range over which the aberrated spot size is a small factor larger than $\delta_x$, meaning the quadratic term is on the order of $\delta_x$,
\begin{equation}\label{FOV_condition}
    \frac{\beta^2 \mathit{FOV}_x^2}{2\delta_x}\propto\delta_x.
\end{equation}
Substituting Eq. \ref{beta} and noting that $w\propto f_\mathit{eff}\lambda/\delta_x$ (in the small angle limit), we arrive at
\begin{equation}\label{FOV_x}
    \mathit{FOV}_x
    \propto 
    \frac{\delta_x^2}{\lambda}\frac{f}{x_m}
    \propto
    \frac{\lambda}{\mathit{NA}^2}\frac{f}{x_m},
\end{equation}
indicating a quadratic relation with the diffraction-limited spot size or the NA, same as the scaling for axial FOV or depth of focus/field (Eq. \ref{FOV_z}). The similarity here stems from the fact that axial geometric aberrations (i.e., defocus) also increases linearly with distance from the focus. 

Note that Eq. \ref{FOV_x} blows up when $x_m\rightarrow 0$, so as mentioned earlier we need to handle this limiting case separately. In the supplemental document, we derive that in the low-NA limit ($w\ll f$) and under certain conditions, the geometric spot size increases \textit{quadratically} with lateral position, $\delta_g(\mathbf{r}_f)\propto\gamma r_f^2$, where
\begin{equation}
    \gamma \propto \frac{w}{f^2}.
\end{equation}
Taylor expanding Eq. \ref{spot_size} to its lowest order greater than 0, we obtain
\begin{equation}\label{taylor2}
    \delta_a(r_f)=\sqrt{\delta_x^2+\gamma^2 r_f^4}\approx
    \delta_x+\frac{\gamma^2 r_f^4}{2\delta_x},
\end{equation}
whereupon, following the same reasoning in Eqs. \ref{FOV_condition} and \ref{FOV_x},
\begin{equation}\label{FOV_x_quad}
    \mathit{FOV}_x
    \propto 
    \delta_x \sqrt{\frac{f}{\lambda}}
    \propto
    \frac{\sqrt{\lambda f}}{\mathit{NA}}.
\end{equation}
Thus, when the entry position is nearly coincident with the optic axis of the parabolic mirror, the FOV scales linearly with the diffraction-limited spot size or the NA.

Another related and important imaging system property is the space-bandwidth product (SBP), or the effective number of independent pixels or resolvable points over the FOV. When $x_m\gg0$, the lateral SBP scales as
\begin{equation}\label{SBP1}
    \mathit{SBP}_\mathit{xy}=\frac{\mathit{FOV}_x\mathit{FOV}_y}{\delta_x\delta_y}\propto \frac{f^2}{\mathit{NA}^2 x_m^2}.
\end{equation}
Interestingly, the lateral SBP at $x_m\gg0$ is invariant to wavelength. However, when $x_m=0$, the lateral SBP scales as
\begin{equation}\label{SBP2}
    \mathit{SBP}_\mathit{xy}\propto \frac{f}{\lambda}.
\end{equation}

In the remaining subsections, we interpret these equations (Eqs. \ref{FOV_x}, \ref{FOV_x_quad}, \ref{SBP1}, and \ref{SBP2}), which were derived for parabolic mirrors under limiting, approximated conditions. We also provide Zemax simulation results on FOV for both parabolic and ellipsoidal mirrors, which agree with these equations when their approximation conditions are valid. Also, notably, the simulations illustrate the gradual transition between the two limiting cases (Eq. \ref{FOV_x} vs. Eq. \ref{FOV_x_quad}).

\subsection{Simulation settings and procedures}\label{sim_settings}
We implemented the system designs in Figs. \ref{fig:parabolic}b and \ref{fig:ellipsoidal}b in Zemax (see Figs. S5 and S6 in the supplement for sample ray diagrams), using ideal paraxial lenses for all lenses ($L_a$, $L_b$, and $L_\mathit{tube}$). For the base parabolic mirror imaging system, we set $f_a=50$ mm, $f_b=50$ mm, $f_\mathit{tube}=300$ mm, $d_t=38$ mm, and $f=19$ mm. For the base ellipsoidal mirror imaging system, we set $f_a=50$ mm, $f_b=50$ mm, $d_t=75.168$ mm, $a=83.5$ mm, and $b=56.480$ mm. These $d_t$ values were chosen to obtain telecentricity at the 90$^\circ$ sample-incidence angle, though we vary this value in our simulations. The mirror dimensions were chosen based on commercially available ellipsoidal (Edmund Optics 90-969) and parabolic (Edmund Optics 68-791) mirrors, which are of similar size to each other.

In our simulations, we systematically varied the multiple design parameters that appear in the above equations (e.g., $\lambda$, $f$, $\mathit{NA}$, etc.) and traced a grid of 45 by 45 fields through the entire imaging system. The beams were apodized so that 99\% of the energy of the Gaussian beam is enclosed within the entrance pupil. We then computed the Strehl ratio and centroids at the conic section mirror focal plane, which was oriented normal to the central chief ray. We used the Airy radius as a proxy for diffraction-limited lateral resolution ($\delta_x$). The grid size was adaptively adjusted according to expected FOV trends from the above derivations for parabolic mirrors (the same trends were used for the ellipsoidal mirror simulations) to efficiently sample the FOVs. We then interpolated the Strehl ratios onto a Cartesian grid, based on the centroids, and estimated $\mathit{FOV}_x$ and $\mathit{FOV}_y$ based on the lateral range over which the Strehl ratio was above 0.8.

While we provide a thorough empirical analysis of FOV of many system configurations in the following sections, we cannot hope to exhaustively capture every possible configuration that may be of interest for the various multi-view imaging applications. Thus, we have made available these Zemax files in Code File 1 (Ref. \cite{code1}) to facilitate further analyses and rapid prototyping (sample Zemax ray diagrams from these files are shown in Figs. S5 and S6 in the supplemental document).

\begin{figure}
    \centering
    \includegraphics[width=.9\columnwidth]{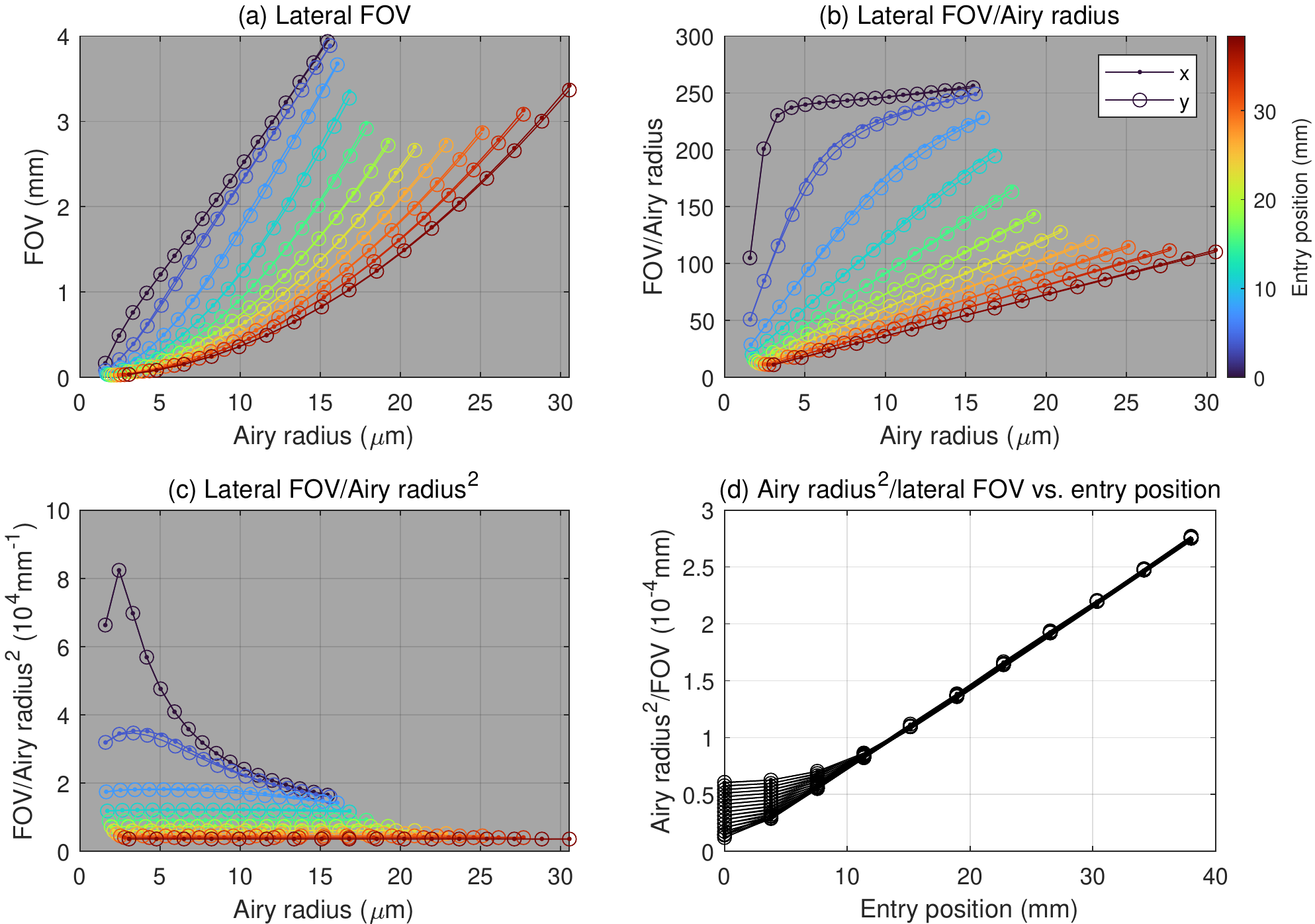}
    \caption{Lateral FOV scaling with lateral resolution at various entry positions ($x_m$) for a parabolic mirror, simulated at $\lambda=800$ nm, $f=19$ mm. There is a gradual shift from a $\mathit{FOV}\propto\delta_x^2$ scaling to a $\mathit{FOV}\propto\delta_x$ scaling as $x_m\rightarrow0$ (a). This gradual shift can be more easily appreciated by normalizing by the Airy radius (b) or its square (c). The results deviate from predictions at higher resolutions ($\delta_x\rightarrow0$), especially when $x_m\rightarrow0$. There is a linear relationship between $x_m$ and 1/FOV when $x_m\gg0$ (d). The different curves in (d) correspond to different Airy radii. Overall, results are nearly identical for $\mathit{FOV}_x$ ($\cdot$) and $\mathit{FOV}_y$ ($\circ$).}
    \label{fig:FOV1_parabolic}
\end{figure}

\subsection{Limitations of theoretical predictions}\label{limitations}
While most of our simulation results follow the derived trends, they deviate when the assumptions we used are violated. In particular, the main assumptions we made were the small-angle approximation, which allowed us to use $w\propto f_\mathit{eff}\lambda/\delta_x$ and $f\gg w$, and that the FOV is small compared to $f$ to allow the Taylor approximations in Eqs. \ref{taylor1} and \ref{taylor2}. The former approximation breaks down when the resolution is very high. On the other hand, the latter approximation breaks down when the resolution is very low, which increases the FOV. In sum, the validity of the simple equations we derived are valid for intermediate resolutions, which will be come clear in the following subsections. Nevertheless, we also simulate beyond the validity conditions to illustrate the effects of doing so.

Another simplifying assumption we made in both the derivation and simulations was that the focal plane is flat. However, for multi-view imaging applications where the goal is to reconstruct across a 3D volume, in principle it doesn't matter if the focal plane is curved. A consequence of this assumption is that FOVs may be underestimated.


\subsection{FOV dependence on NA or lateral resolution and $x_m$}
The simulations in Fig. \ref{fig:FOV1_parabolic}a-c verify the linear and quadratic dependence of lateral FOV on $\mathit{NA}$ for parabolic mirrors when $x_m\gg0$ (Eq. \ref{FOV_x}) and $x_m=0$ (Eq. \ref{FOV_x_quad}), respectively. Interestingly, these trends also extend to ellipsoidal mirrors (Fig. S1a-c). These simulations also show the gradual transition from quadratic to linear dependence of both $\mathit{FOV}_x$ and $\mathit{FOV}_y$ on $\mathit{NA}$ or $\delta_x$ as $x_m\rightarrow0$. Furthermore, when still in the quadratic regime, the parabolic mirror simulations show that FOV scales like $1/x_m$ (Fig. \ref{fig:FOV1_parabolic}d), consistent with our derivation (Eq. \ref{FOV_x_quad}). For ellipsoidal mirrors, this trend is also quite linear (perhaps slightly super-linear; Fig. S1d). However, note that while $x_m$ is theoretically unbounded for parabolic mirrors, it is restricted to a maximum value of $b$ for ellipoidal mirrors. Finally, we note that when $x_m=0$ and $\delta_x\rightarrow0$, the simulations deviate from the predicted linear relationship with $\delta_x$, which is not too surprising, given that the small angle approximation breaks down, as discussed in Sec. \ref{limitations}.

The fact that axial and lateral FOVs have the same scaling (Eqs. \ref{FOV_z} and \ref{FOV_x}) when $x_m\gg0$ is very fortuitous, as that means the 3D FOV remains roughly isotropic and therefore no information is wasted when computationally combining multi-view images. For example, if the lateral FOV were much larger than the axial FOV, then the information at the lateral periphery would be wasted, as it would fall outside of the depth of field from a perpendicular view. Put another, more optimistic way, for multi-angle imaging applications, the fact that conic section mirrors have limited lateral FOVs is \textit{not} actually a limitation, even at high NAs, because of the concomitantly limited depth of field. Even when $x_m\rightarrow0$, it is evident from Figs. \ref{fig:FOV1_parabolic} and S1 that the FOV only increases for any given NA.

\begin{figure}
    \centering
    \includegraphics[width=.9\columnwidth]{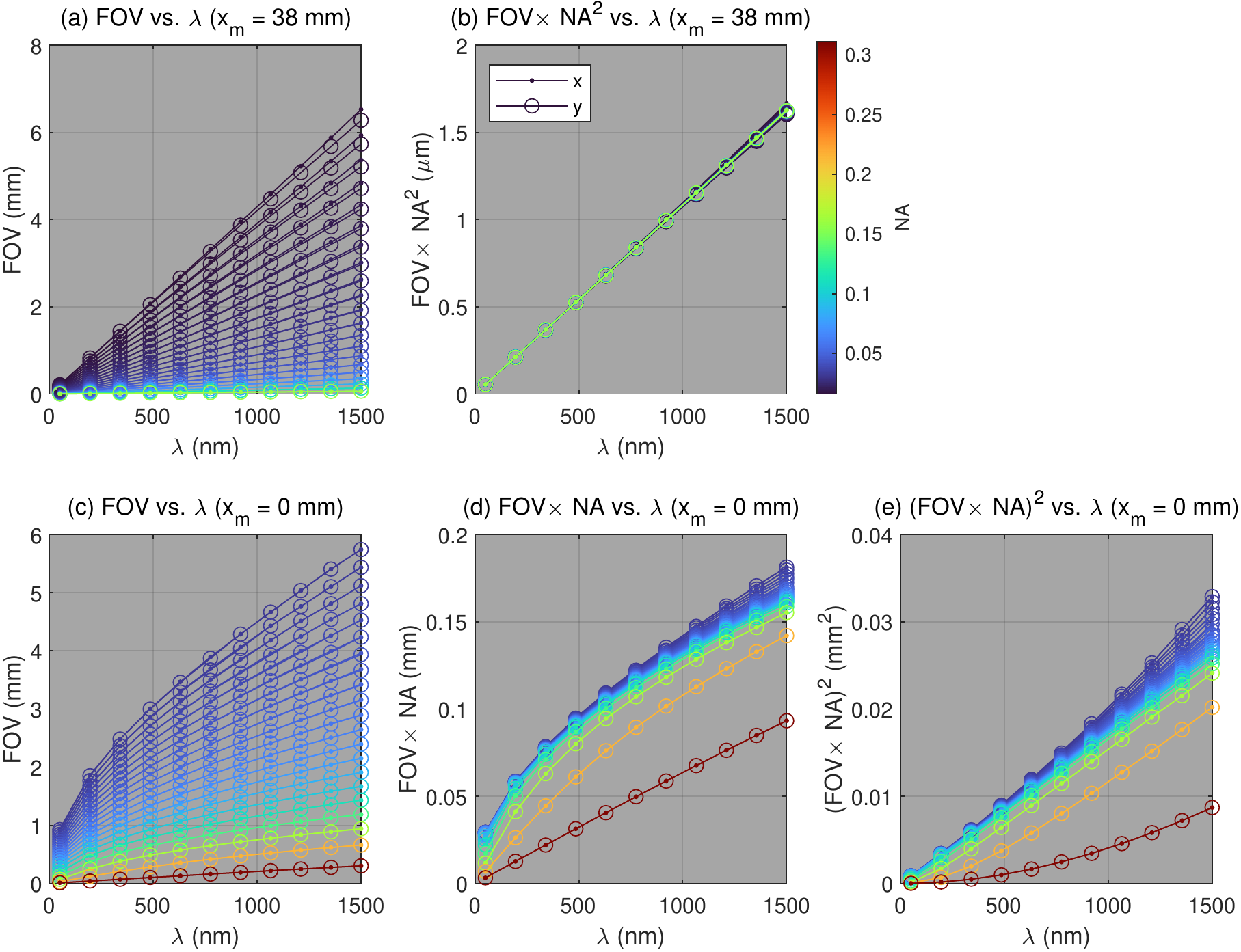}
    \caption{Lateral FOV scaling with wavelength ($\lambda$) for a parabolic mirror, simulated at $f=19$ mm and two entry positions, $x_m=38$ mm (first row) and $x_m=0$ mm (second row). At $x_m=38$ mm, the FOV scales linearly with $\lambda$ (a). Once we account for the $\mathit{NA}^2$ dependence verified in Fig. \ref{fig:FOV1_parabolic}, all the curves collapse into a single line that reflects pure $\lambda$ dependence (b). However, at $x_m=0$, FOV is proportional to $\sqrt{\lambda}$ (c). Accounting for the $\mathit{NA}^2$ dependence partially collapses the curves (d), and the approximately linear relation after squaring (e) confirms the square-root relation in (c), consistent with Eq. \ref{FOV_x_quad}. Note that the curves deviate from the model at both high and low NAs, where the assumptions of the derivations are violated (Sec. \ref{limitations}). Note also that $\mathit{FOV}_x$ ($\cdot$) and $\mathit{FOV}_y$ ($\circ$) are nearly identical. }
    \label{fig:FOV2_parabolic}
\end{figure}


\subsection{FOV dependence on wavelength}

Eq. \ref{FOV_x} predicts a linear increase in FOV with wavelength when $x_m\gg0$, while Eq. \ref{FOV_x_quad} predicts a square-root increase in FOV when $x_m=0$. These trends are confirmed in simulation for both parabolic (Fig. \ref{fig:FOV2_parabolic}) and ellipsoidal (Fig. S2) mirrors. Interestingly, for the case of $x_m\gg0$, the linear trend holds for a wide range of NAs (Figs. \ref{fig:FOV2_parabolic}a-b and S2a-b), while for $x_m=0$, the results slightly deviate from the square-root trend at higher and lower NAs (more so for the former; Figs. \ref{fig:FOV2_parabolic}d-e and S2d-e). Since the diffraction-limited spot size also increases linearly with $\lambda$, for $x_m\gg0$ the SBP is invariant to $\lambda$ (Eq. \ref{SBP1}). However, for $x_m=0$, the FOV increases more slowly than the spot size, so that the SBP actually gets worse as $\lambda$ increases (Eq. \ref{SBP2}).


\begin{figure}
    \centering
    \includegraphics[width=.9\columnwidth]{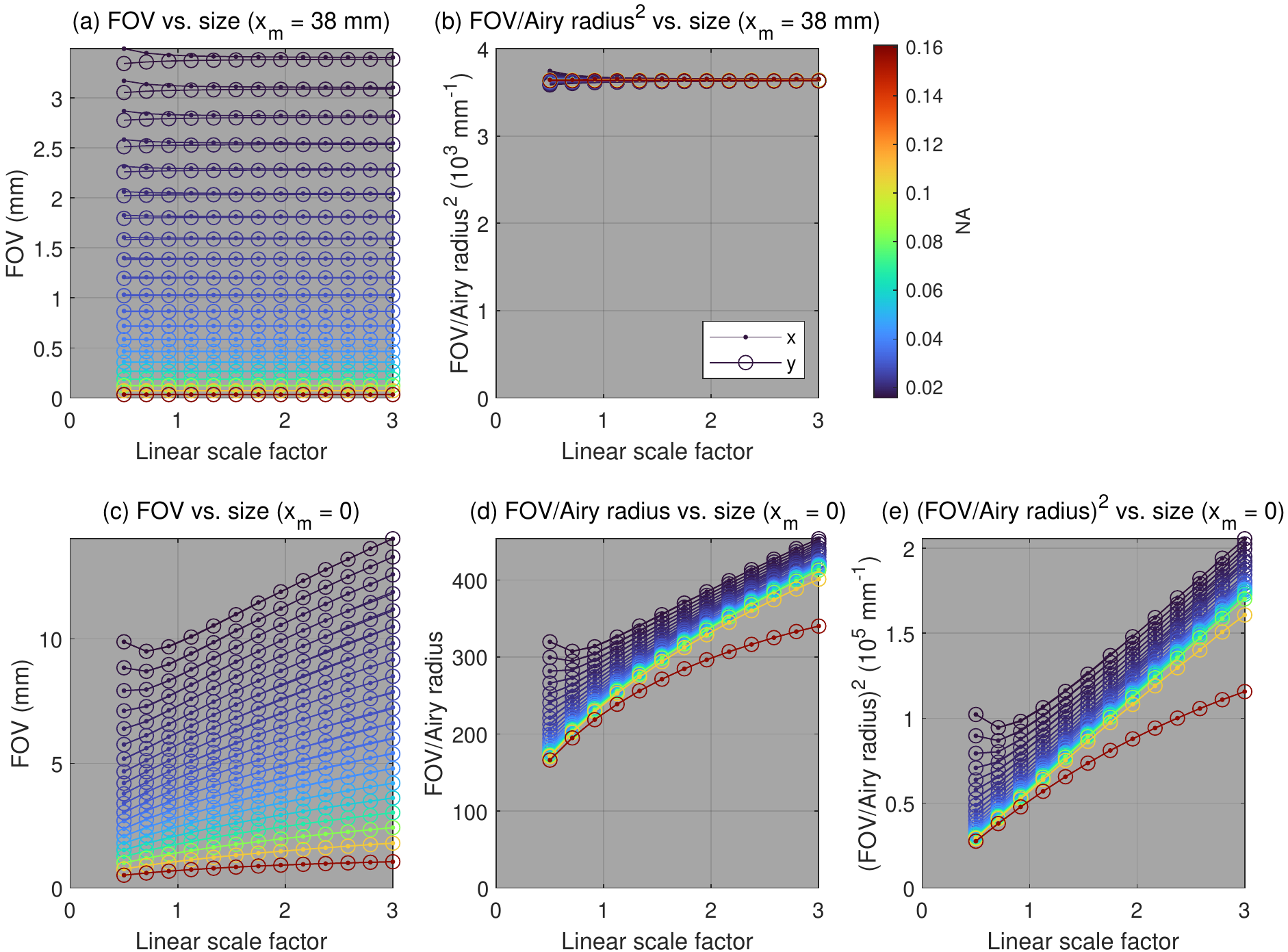}
    \caption{Lateral FOV scaling with parabolic mirror size. Here, a linear scale factor of 1 corresponds to $f=19$ mm, simulated at $\lambda=800$ nm. When $x_m=38$ mm, the FOV is approximately invariant to linear scaling (a-b). However, when $x_m=0$ mm, the FOV scales by approximately the square-root of the mirror size (c), which is easier to appreciate when removing the $\delta_x$ dependence (d) and squaring (e). Results for $\mathit{FOV}_x$ ($\cdot$) and $\mathit{FOV}_y$ ($\circ$) are nearly identical.}
    \label{fig:FOV3_parabolic}
\end{figure}


\subsection{FOV dependence on isotropic scaling of mirror size}\label{size}
The scale of parabolic mirrors can be tuned via its focal length $f$, while the scale of ellipsoidal mirrors can be tuned by scaling its semi-major and semi-minor axes, $a$ and $b$, by the same factor (we did not investigate aspect ratio, which is left for future studies). For an aberration-free imaging system, the FOV would increase proportionally if all the linear dimensions were isotropically expanded (and thus the SBP would increase quadratically) \cite{lohmann1989scaling}. However, the tilt aberrations prevent conic section mirrors from enjoying this property. In fact, when $x_m\gg0$, the FOV and SBP are invariant to isotropic scaling of mirror size, due to the linear increase in geometric aberrations as a function of distance from the nominal focus (Eq. \ref{beta}). Incidentally, this invariance is also true of the axial dimension, in which geometric aberrations (defocus) also increases linearly with distance from the nominal focus. In particular, Eq. \ref{FOV_z} does not depend on the size of the focusing optics. Although Eq. \ref{FOV_x} does appear to depend on the size of the mirror, which is dictated by $f$, note that $f/x_m$ is fixed -- that is, if $f$ is doubled, then $x_m$ must also be doubled to maintain the same sample-incident inclination angle (Eq. \ref{para_theta}). However, when $x_m=0$, we expect FOV to increase with the square-root of the scale of mirror (Eq. \ref{FOV_x_quad}), and the SBP to increase linearly with the scale (Eq. \ref{SBP2}). The simulations support this trend not only for parabolic mirrors (Fig. \ref{fig:FOV3_parabolic}), but also for ellipsoidal mirrors (Fig. S3). 

Thus, increasing the size of conic sections may not be an effective strategy for increasing the FOV or SBP, unless the imaging system operates close to $x_m=0$. Nevertheless, the scale-invariance for $x_m\gg0$ can still be practically useful as it allows system miniaturization without sacrificing performance.

\begin{figure}
    \centering
    \includegraphics[width=.9\columnwidth]{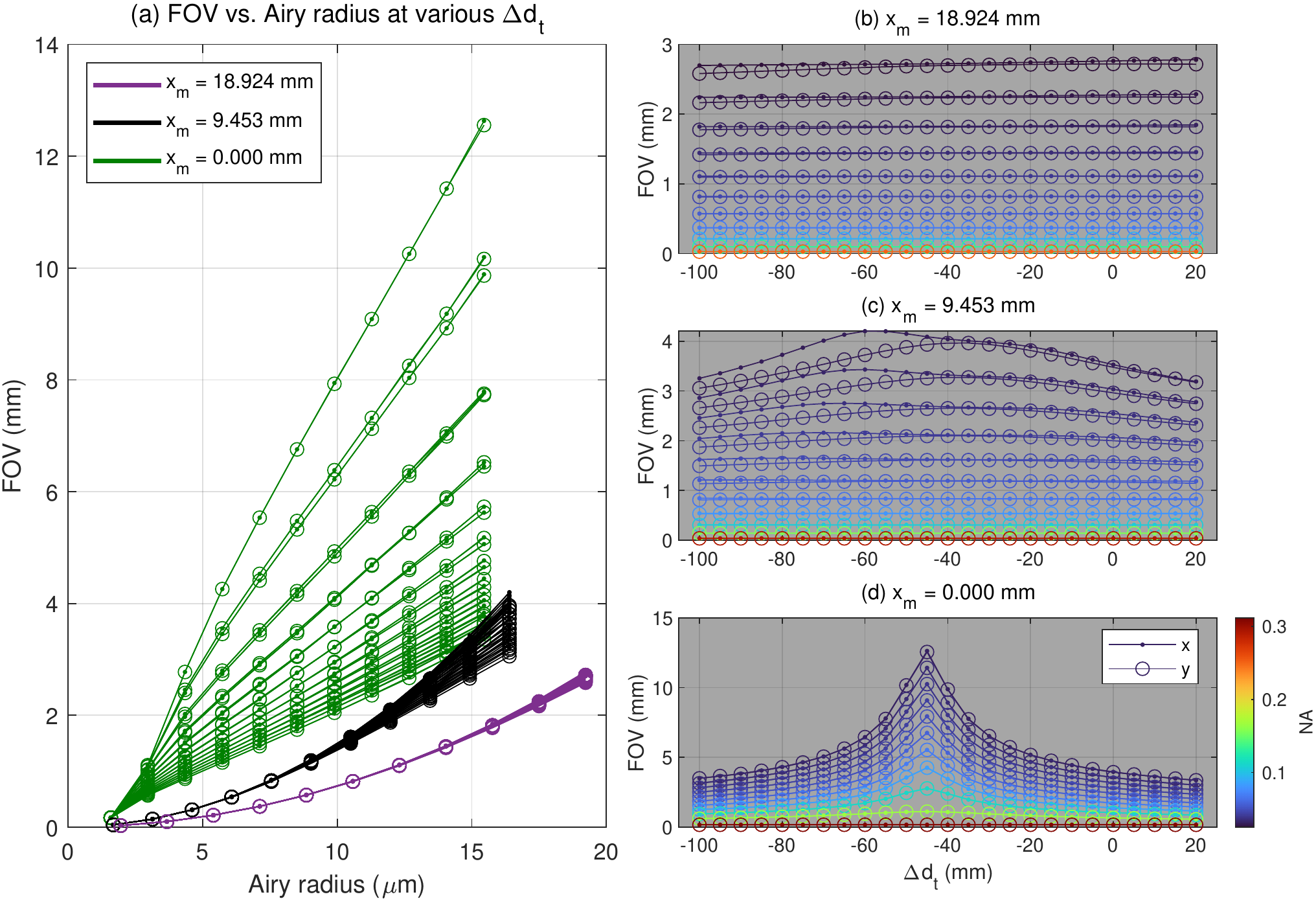}
    \caption{Effect telecentricity on lateral FOV for a parabolic mirror, simulated at $\lambda=800$ nm and $f=19$ mm. In (a), each color corresponds to a different $x_m$ and the individual curves within correspond to different $\Delta d_t$ values, which confer varying degrees of non-telecentricity. When $\Delta d_t=0$, the 90$^\circ$-sample-incidence configuration is telecentric. FOV sensitivity to $d_t$ is greatest at $x_m=0$ and decreases as $x_m$ increases. The same data is plotted in (b-d) against $\Delta d_t$, where now it is clear that the FOV attains a maximum when $\Delta d_t\approx-45$ mm when $x_m=0$ (d). Results for $\mathit{FOV}_x$ ($\cdot$) and $\mathit{FOV}_y$ ($\circ$) are nearly identical, except at very low NAs when $x_m>0$ (b-c).}
    \label{fig:FOV4_parabolic}
\end{figure}

\subsection{FOV dependence on telecentricity}\label{FOV_telecentricity}
We can understand the effects of (non-)telecentricity through simulation by varying $d_t$ by $\Delta d_t$ (Figs. \ref{fig:FOV4_parabolic} and S4), in the absence of theoretical predictions. When $\Delta d_t=0$, the telecentric status defaults to the base case defined in Sec. \ref{sim_settings}. For both parabolic (Fig. \ref{fig:FOV4_parabolic}) and ellipsoidal (Fig. S4) mirrors, when $x_m\gg0$, the FOV is invariant to whether the rays are telecentric. This property adds flexibility to system designs and alignment. However, the closer $x_m$ approaches 0, the more the degree of non-telecentricity matters. In fact, the FOV attains a maximum value for $\Delta d_t\approx-45$ mm for our simulated parabolic mirror (Fig. \ref{fig:FOV4_parabolic}d) and $\Delta d_t\approx-5.8$ mm for our simulated ellipsoidal mirror (Fig. S4d). These configurations are slightly non-telecentric and appear to maximize the focal plane flatness. We note that when $\Delta d_t$ deviates from these values, the lateral FOV may still be large along a curved focal plane, as previously discussed (Sec. \ref{limitations}).



\begin{figure}
    \centering
    \includegraphics[width=.6\textwidth]{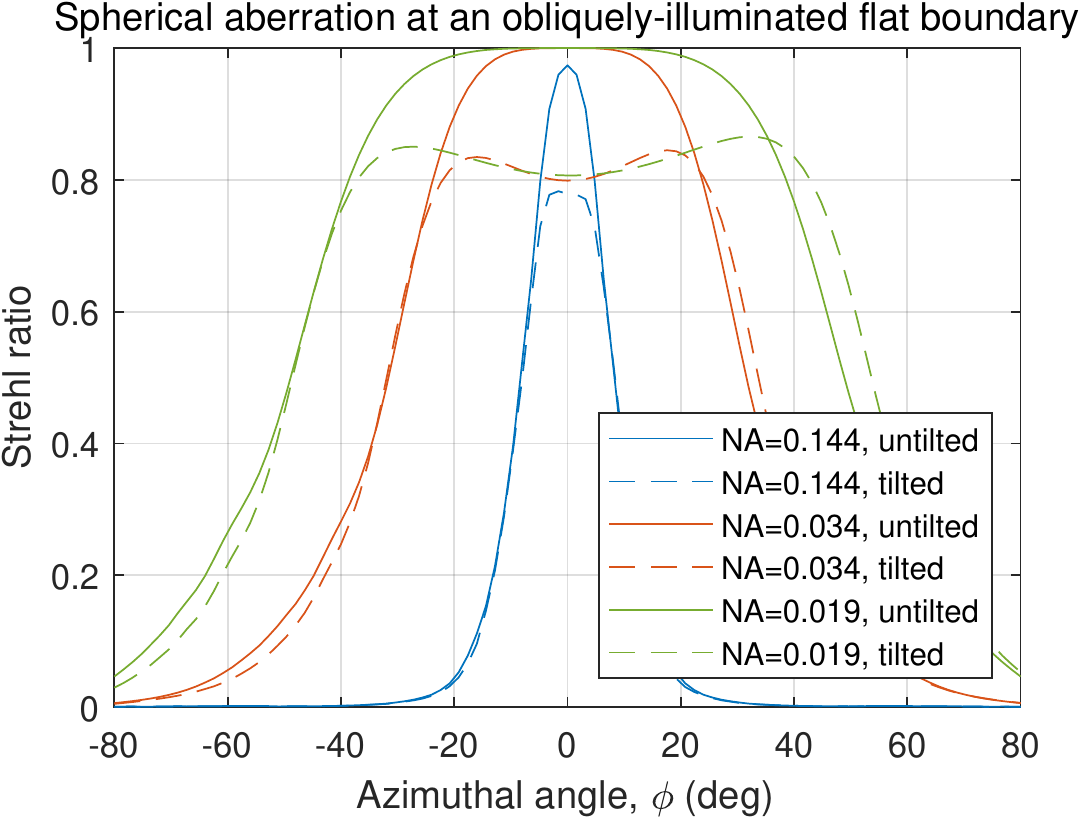}
    \caption{Simulations of the quality of beam focusing by a parabolic mirror as a function of the nominal azimuthal angle $\phi$ of incidence to the sample, where the beam has to traverse a flat 1.5-mm-thick glass (BK7) slide followed by water (Fig. \ref{fig:spherical_aberration_correction}a). The bottom surface is the glass slide 6.5 mm above the nominal focus of the parabolic mirror. Focus quality is assessed using the Strehl ratio at the shifted focal plane for three different NAs (0.019, 0.034, 0.144) and two different beams: an untilted (center of the imaging FOV) and tilted beam (edge of the FOV). The tilted beam was chosen so that in air, the Strehl ratio is $\sim$0.8. The larger the NA, the smaller the angular range over which good focusing performance is achievable.}
    \label{fig:spherical_aberrations}
\end{figure}

\section{Compensating for spherical aberration at obliquely-illuminated flat boundaries}\label{spherical_aberrations}
\begin{figure}
    \centering
    \includegraphics[width=.95\textwidth]{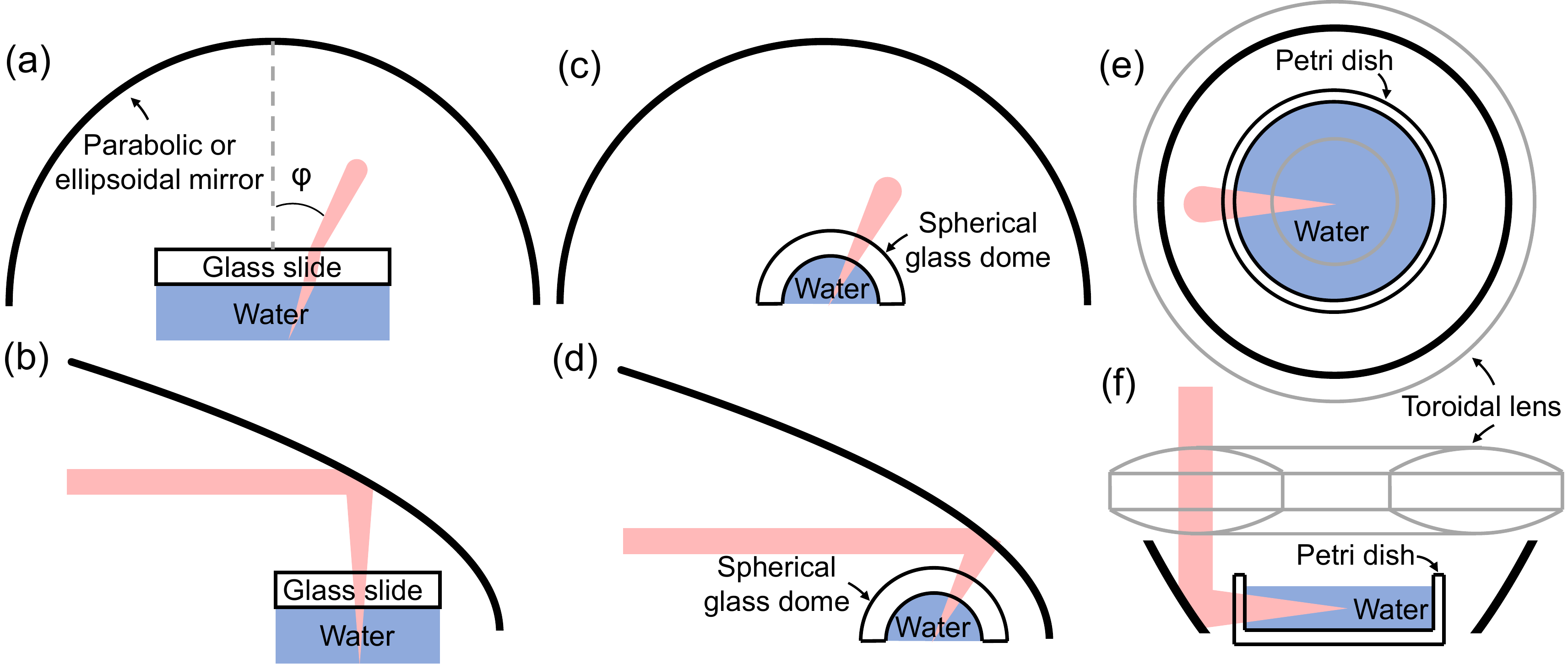}
    \caption{Proposed strategies for reducing spherical aberrations at obliquely illuminated flat boundaries, depicted in (a-b). The first row views the imaging configurations along the optic axis of the mirror, while the second row along an orthogonal direction. One strategy is to use a spherical optical dome (c-d), which eliminates oblique illumination angles. Another strategy (for the single-axis 360$^\circ$ configuration) is to use the combination of a toroidal lens and a Petri dish or tube (e-f).}
    \label{fig:spherical_aberration_correction}
\end{figure}

Depending on the application and the way the sample is mounted, the angularly varying incident beam may encounter interfaces that can cause aberrations. For example, spherical aberrations can be serious if there is a flat refractive index (RI) discontinuity before the sample (or at the surface of the sample itself). In this situation, although near-normal incidence angles may be negligibly aberrated if the NA is sufficiently small, the more oblique the incidence angle and the higher the NA, the worse the aberrations (Figs. \ref{fig:spherical_aberrations} and \ref{fig:spherical_aberration_correction}a). Another disadvantage is that the maximum sample-incident angle becomes more restricted due to Snell's law (e.g., when switching from medium 1 to 2, the maximum angle in medium 2 is $\sin^{-1}(n_1/n_2)$). This can be problematic for techniques that aim to reconstruct sub-surface information (e.g., OPT, ODT, OCPT, OCRT) if the sample itself exhibits a flat surface or if it is submerged in water with flat air-water or air-glass interface. 

A workaround is to fill the entire mirror with water or another RI-matching medium to eliminate all RI boundaries, akin to water- and oil-immersion objectives. Such an approach may be practical if the mirror were shrunken, which does not affect the theoretical FOV for certain inclination angles (Sec. \ref{size}). Here, we propose two additional solutions that introduce additional optical elements to overcome these spherical aberrations at oblique illumination angles: 1) an optical dome (Fig. \ref{fig:spherical_aberration_correction}c-d), and 2) a cylindrical Petri dish or tube with a toroidal lens (Fig. \ref{fig:spherical_aberration_correction}e-f).

\subsection{Optical dome}\label{dome}
One way to substantially reduce spherical aberrations is to use an optical dome, a spherical shell lens that can be parameterized by two concentric spherical surfaces with radii $r_\mathit{outer}>r_\mathit{inner}$ and uniform thickness $r_\mathit{outer}-r_\mathit{inner}$. Domes have the interesting property that a point source placed at its center (i.e., the center of its concentric spherical surfaces) is unaffected by its presence, apart from a constant path delay, because all rays are perpendicular to the dome's two surfaces (equivalently, the wavefront curvature matches that of the surfaces). Thus, one strategy to avoid spherical aberrations is to align a dome's center with a conic section mirror's focus, so that the sample-incident rays are never oblique to the air-glass or air-immersion-medium interface. The space between the dome and the sample can be filled with the immersion medium to avoid oblique incidence angles at the sample surface. An optical dome can be used for either the single-axis 360$^\circ$ configuration or the two-axis configuration (Fig. \ref{fig:spherical_aberration_correction}c-d).

To demonstrate the effectiveness of optical domes, we performed a similar simulation to that in Fig. \ref{fig:spherical_aberrations}, where we used an optical dome with $r_\mathit{outer}=8$ mm and $r_\mathit{inner}=6.5$ mm, centered at a parabolic mirror's nominal focus. We simulated both an untilted beam (i.e., one which would focus to the center of the FOV, the mirror's nominal focus), which in air would achieve a Strehl ratio of 1.0, and a tilted beam (i.e., one which would focus off-center) with a tilt angle that would achieve a Strehl ratio of $\sim$0.8 in air. We repeated this simulation for three different NAs (0.019, 0.034, and 0.144), and in all cases, the Strehl ratio of the untilted beam was maintained at 1.0 and that of the tilted beam remained at $\sim$0.8. Due to the symmetry of both the dome and the mirror, focusing performance is independent of azimuthal angle and thus preserves the wide-angle multi-view imaging capabilities of conic section mirrors. We note that optical domes can be used for other multi-view imaging strategies besides conic section mirrors.


\subsection{Cylindrical Petri dish or tube with a toroidal lens}\label{petri}
Another strategy to avoid oblique incidence angles for the single-axis 360$^\circ$ configuration is to use a cylindrical Petri dish or tube, which perhaps can be thought of as a 1D version of an optical dome. This element is essentially a tube with radii $r_\mathit{outer}>r_\mathit{inner}$ and a uniform thickness. However, using this element alone would still result in an astigmatic focus, since it only has matching curvature in one dimension. To correct this astigmatism, we can introduce a toroidal lens before the mirror, which adds additional focusing power in the Petri dish's uncurved dimension (Fig. \ref{fig:spherical_aberration_correction}e-f), in the same way that the axicon mirror that normally can only focus in 1D can focus in 2D with the help of a toroidal lens \cite{mcnabb2015complete}. While there is some flexibility in the design of the toroidal lens, for illustrative purposes we have included sample toroidal lens designs in the accompanying Zemax files for both parabolic and ellipsoidal mirrors (Code 1 \cite{code1}; see also Figs. S5 and S6). Using the parabolic model, we performed the same simulation as we did for an optical dome (Sec. \ref{dome}) and flat glass slide boundary (Fig. \ref{fig:spherical_aberrations}), using a 35-mm-diameter Petri dish (BK7) with a 1-mm wall thickness (see Zemax file for toroidal lens design \cite{code1}). For the two lower NAs out of 0.019, 0.034, and 0.144, the untilted Strehl ratios were nearly 1.0 (>0.998) and the tilted Strehl ratios were >0.84. For the highest NA simulated (0.144), the untilted and tilted Strehl ratios were <0.4. Hence, for this strategy to work for higher NAs, more sophisticated toroidal lens design may be necessary. However, for lower NAs, which are necessary for larger FOVs (both axially and laterally), this Petri-dish-toroidal-lens combination is a good strategy and convenient for mounting biological samples.

\section{Other practical considerations}
\label{illumination}

\subsection{In-plane rotation of field of view}
A peculiarity of these conic section mirrors is that when the entry position is scanned across the mirror aperture such that the azimuthal incidence angle changes, the FOV of the sample rotates around the central chief ray for that particular azimuthal angle. This can either be solved in software by digitally rotating the images, or in hardware, if using lateral point-scanning, by rotating the raster scan pattern together with the azimuthal incidence angle so that there is a spin-orbit lock (cf., the synchronous rotation of the Moon).

\subsection{Point-, line-, and full-field illumination and detection}\label{camera}
As we mentioned earlier, although our descriptions were primarily from the perspective of point-scanning systems, our results are also applicable to camera-based systems. In particular, instead of the first scanning mirror in collimated space, which performs lateral scanning, we can instead put a 2D camera in the preceding anticonjugate plane (the leftmost planes in Figs. \ref{fig:parabolic}b and \ref{fig:ellipsoidal}b). As such, the chief and marginal rays of the camera-based system and point-scanning system are identical when the stop of the latter is restricted by the scanning mirror extent. 
For generating multi-angle, wide-field, plane-wave illumination, as required by common implementations of imaging techniques such as ODT, FPM, and OPT, one would need to anticonjugate the illumination relative to the point-scanning configuration. That is, in all planes where the beam would be focusing for a point-scanning system, the beam should instead be collimated and vice versa. For example, a parabolic mirror-based wide-field illumination scheme was recently reported for FPM \cite{lee2019reflective}. Line-field (1D) illumination and detection with 1D lateral scanning is essentially a hybrid, behaving like a point-scanning system in one lateral dimension and a full-field/camera-based system in the other.

\subsection{Transmissive vs. reflective imaging}\label{trans_refl}
While for many applications, particularly those in reflective imaging geometries, one imaging path is sufficient, for some transmissive geometries a separate but similar or identical imaging path is desired. For example, in OPT, there is an illumination path that generates collimated light incident on the sample, and an imaging path that images the transmitted light onto a detector. In such situations, a full rotationally symmetric conic section mirror can be used, such that 180$^\circ$ is used for illumination and the other 180$^\circ$ is used for imaging.


\subsection{Inclination-angle-dependent resolution, magnification, and telecentricity}
If the radial entry position across the mirror aperture is varied to change the sample-incident inclination angle, the effective focal length changes while the input beam size remains the same. As a result, the lateral resolution, magnification, and telecentricity also change. Zoom systems or lenses with dynamically tunable foci may be used before the mirror to compensate for such changes. For a galvanometric point-scanning based imaging system, the change in magnification can be compensated by dynamically adjusting the lateral scan range. For a multi-camera system, each individual camera and its associated lenses may be separately designed according to the conic section mirror's effective focal length.

\subsection{Calibration}
The theoretical analyses in this paper assume ideal conditions -- perfect alignment, perfectly parabolic or ellipsoidal mirror shapes, and aberration-free auxiliary optical elements (e.g, tube lenses). Since such conditions are impossible to meet in practice, it may be necessary to precalibrate the spatioangular scan parameters for all view angles, especially for computational imaging approaches that perform, for example, 3D reconstructions. One such calibration procedure would be to acquire multi-view images of a reference 3D target with well-defined features, and to jointly register these features and the multi-view calibration parameters in a manner similar to how photogrammetry/SfM jointly estimate camera poses and 3D object point clouds \cite{ullman1979interpretation, wu2013towards,furukawa2015multi,schonberger2016structure,zhou2021mesoscopic}. The calibration procedure would be further simplified if the positions 3D features of the target are known.







\section{Discussion and conclusion}
We have presented a general set of strategies employing conic section mirrors, particularly parabolic and ellipsoidal mirrors, for performing multi-angle imaging over very wide angular ranges over one or two axes without requiring sample rotation. Thus, high-speed 3D imaging is possible through fast optomechanical scanning or multiple cameras. Through derivations and comprehensive, but not exhaustive, simulations, we have established key scaling relationships between FOV/SBP and imaging system parameters, notably the resolution or NA, wavelength, mirror size, effective focal length, and telecentricity (Sec. \ref{fov}; Eqs. \ref{FOV_x}, \ref{FOV_x_quad}). Importantly, we have argued that for multi-angle 3D imaging applications, the well-known tilt aberrations of parabolic and ellipsoidal mirrors do not restrict FOV any more than depth of field limits axial FOV. In particular, the scaling is identical (i.e., quadratic) for most practical sample-incident inclination angles. Interestingly, the FOV dependencies are similar for both parabolic and ellipsoidal mirrors, and thus both are in theory viable options for multi-view imaging applications.

We have left out discussions on other conic section mirrors, such as rotationally symmetric hyperbolic (hyperboloidal), spherical, toroidal, and axicon mirrors, which could also be able to achieve single-axis 360$^{\circ}$ imaging performance when combined with other elements. For example, an axicon mirror could be combined with a toroidal lens in a similar manner discussed in Sec. \ref{petri} to provide focusing power in the unfocused dimension (e.g., see Fig. 3 of \cite{mcnabb2015complete}).

Our proposed approaches allow for very wide angular ranges, which would lead to spherical aberrations at obliquely illuminated flat RI discontinuities (e.g., glass coverslip) if uncorrected. We have proposed multiple strategies for compensating for these aberrations (Sec. \ref{spherical_aberrations}). Note that this problem is not intrinsic to our approaches, but rather a general one for high-angle imaging applications -- this is why some high-NA objectives have correction collars for glass coverslips. 

As many imaging techniques benefit from imaging over wide angular ranges, the applicability and generality of our approach is broad. For example, a symmetric parabolic or ellipsoidal mirror can be used to collect 2D transmissive projection images over 180$^\circ$ for OPT or OCPT. Similarly, these approaches can be applied to epi-mode multi-angle imaging techniques such as OCRT and photogrammetry. 

We hope that our theoretical analyses of conic section mirrors will prove useful to researchers designing multi-angle imaging systems. To this end, we have provided Zemax files (Code 1 \cite{code1}) as generic templates for parabolic- and ellipsoidal-mirror-based system designs, including configurations for spherical aberration compensation, to facilitate adoption.

\medskip
\noindent
\textbf{Funding.} National Science Foundation (CBET-1902904).

\medskip
\noindent
\textbf{Disclosures.} KCZ, AHD, RPM, RQ, SF, JAI: Duke University (P). 

\medskip
\noindent
\textbf{Data availability.} Data underlying the simulation results presented in this paper can be generated from Code File 1 (Ref. \cite{code1}).

\medskip
\noindent
\textbf{Supplemental document.} See Supplement 1 for supporting content.

\bibliography{sample}

\end{document}


\maketitle

\section{Derivation of geometric spot size for parabolic mirrors}
To determine $\delta_g(\mathbf{r}_f)$ (see Eq. 20) for a parabolic mirror, we analytically trace rays, parameterized by direction $\mathbf{u}=(u_x,u_y)$ and the position at which the ray intersects the mirror surface $\mathbf{r}_m=(x_m,y_m)$. Note that $u_z=-\sqrt{1-u_x^2-u_y^2}$ and $z_m=P\left(\sqrt{x_m^2+y_m^2}\right)-f$, where we have subtracted out the focal length from the z position so that the focus is at the origin. All derivations in this section assume non-telecentric lateral scanning, with the input collimated beam pivoting about the mirror surface. This scanning configuration significantly simplifies the math, but note the caveat at the end of this section.

The new ray direction upon reflection off of the mirror is given by
\begin{equation}\label{reflection}
    \mathbf{u}_m(\mathbf{u},\mathbf{r}_m)=\mathbf{u}-2[\mathbf{u}\cdot\hat{n}(\mathbf{r}_m)]\hat{n}(\mathbf{r}_m),
\end{equation}
where $\hat{n}(\mathbf{r}_m)=\left(-x_m,-y_m,2f\right)/\sqrt{x_m^2+y_m^2+4f^2}$ is the mirror surface normal unit vector. We then propagate this ray from the mirror to the focal plane, which is defined as the plane intersecting the focus (the origin) with a surface normal, oriented according to the central chief ray for a given parabolic mirror aperture entry position, $\mathbf{r}_m$:
\begin{equation}
\hat{n}_f(\mathbf{r}_m)=\mathbf{u}_m(\mathbf{0},\mathbf{r}_m).
\end{equation}
The distance an arbitrary ray, defined by direction $\mathbf{u}$ and mirror intersection point $\mathbf{r}_m'$, must travel from the mirror to the focal plane is given by
\begin{equation}
    d_\mathit{f}(\mathbf{u},\mathbf{r}_m,\mathbf{r}_m')=\frac{-\hat{n}_f(\mathbf{r}_m)\cdot\mathbf{r}_m'}{\hat{n}_f(\mathbf{r}_m)\cdot\mathbf{u}_m(\mathbf{u}',\mathbf{r}_m')}.
\end{equation}
Thus, the 3D position of the arbitrary ray at the focal plane for a given central chief ray is given by
\begin{equation}\label{final_ray}
    \mathbf{r}_f(\mathbf{u},\mathbf{r}_m,\mathbf{r}_m')=
    \mathbf{r}_m'+
    d_\mathit{f}(\mathbf{u},\mathbf{r}_m,\mathbf{r}_m')\mathbf{u}_m(\mathbf{u},\mathbf{r}_m').
\end{equation}

Recall that the goal here is to understand how the geometric spot size varies as a function of lateral position ($\delta_g(\mathbf{r}_f)$ in Eq. 20) and potentially the diffraction-limited spot size ($\delta_x$). 
However, as we confirm with simulation (Sec. 5), such dependencies themselves depend on the entry position, $x_m$. We consider two limiting cases: 1) $x_m\gg0$ (off-axis) and 2) $x_m=0$ (on-axis). 

\subsection*{Geometric spot size when $x_m\gg0$}
Let's start with the first case by performing a first-order Taylor expansion of Eq. \ref{final_ray} at $(u_x=0,u_y=0)$, which we denote as as $\widetilde{\mathbf{r}}_f^{(1)}(\mathbf{u},\mathbf{r}_m,\mathbf{r}_m')$; that is,
\begin{equation}\label{ray_1st_order}
    \mathbf{r}_f(\mathbf{u},\mathbf{r}_m,\mathbf{r}_m')=
    \widetilde{\mathbf{r}}_f^{(1)}(\mathbf{u},\mathbf{r}_m,\mathbf{r}_m')
    +
    O\left((u_x+u_y)^2\right),
\end{equation}
where the superscript $(1)$ denotes that it is a first-order approximation. This expansion is justified by the relatively narrow range of lateral scanning relative to the mirror dimensions. Since Eqs. \ref{final_ray} and \ref{ray_1st_order} describe where an arbitrary 3D ray trace terminates, we can propagate rays comprising a collimated input beam of radius, $w/2$, and estimate the geometric spot width at the focal plane. Since even the Taylor approximation (Eq. \ref{ray_1st_order}) is a complicated expression (not shown here), we further simplify the lateral spot size analysis by exploiting the rotational symmetry of parabolic mirrors by setting $y_m=0$ and analyzing four simple 1D cases that analyze $x$ and $y$ separately for both position and spot size. In particular, the four cases are
\begin{enumerate}
    \item laterally scanning \underline{$x$} by setting \underline{$u_y=0$} and varying only \underline{$u_x$}, and analyzing the 1D geometric spot size in \underline{$x$} by propagating two marginal rays at entry positions \underline{$(x_m'=x_m+w/2,y_m'=0)$} and \underline{$(x_m'=x_m-w/2,y_m'=0)$}:
    \begin{equation}\label{case1}
    \begin{aligned}
        \delta_{g,x}(u_x,x_m)\propto
        &\hphantom{A}  
        \big|
        \widetilde{\mathbf{r}}_f^{(1)}\left(\mathbf{u}=(u_x,0),\mathbf{r}_m=\left(x_m,0\right),\mathbf{r}_m'=\left(x_m+w/2,0\right)\right)-\\
        &\hphantom{A}\hphantom{\big|}  
        \widetilde{\mathbf{r}}_f^{(1)}\left(\mathbf{u}=(u_x,0),\mathbf{r}_m=\left(x_m,0\right),\mathbf{r}_m'=\left(x_m-w/2,0\right)\right)
        \big|\\
        \propto&
        \hphantom{A}  
        x_mf_\mathit{eff}u_x\frac{w}{2}
        \frac{\bigg|
        \left(\frac{w}{2}\right)^4(-12f^2+x_m^2)-
        32\left(\frac{w}{2}\right)^2f^2f_\mathit{eff}^2+
        64f^3f_\mathit{eff}^3
        \bigg|}
        {
        \bigg|
        \left(\frac{w}{2}\right)^4(-4f^2+x_m^2)^2-
        128\left(\frac{w}{2}\right)^2f^3f_\mathit{eff}^3+
        256f^4f_\mathit{eff}^4
        \bigg|
        }\\
        \approx&
        \hphantom{A}  
        \frac{x_m u_x w}{8f},
    \end{aligned}
    \end{equation}
    where the approximation assumes $f\gg w$.

    \item laterally scanning \underline{$y$} by setting \underline{$u_x=0$} and varying only \underline{$u_y$}, and analyzing the 1D geometric spot size in \underline{$y$} by propagating two marginal rays at entry positions \underline{$(x_m'=x_m,y_m'=w/2)$} and \underline{$(x_m'=x_m,y_m'=-w/2)$}:
    \begin{equation}
    \begin{aligned}
        \delta_{g,y}(u_y,x_m)\propto
        &\hphantom{A}  
        \big|
        \widetilde{\mathbf{r}}_f^{(1)}\left(\mathbf{u}=(0,u_y),\mathbf{r}_m=\left(x_m,0\right),\mathbf{r}_m'=\left(x_m,w/2\right)\right)-\\
        &\hphantom{A}\hphantom{\big|}  
        \widetilde{\mathbf{r}}_f^{(1)}\left(\mathbf{u}=(0,u_y),\mathbf{r}_m=\left(x_m,0\right),\mathbf{r}_m'=\left(x_m,-w/2\right)\right)
        \big|\\
        \propto&
        \hphantom{A}  
        x_mf_\mathit{eff}u_x\frac{w}{2}
        \frac{
        \left(\frac{w}{2}\right)^2 + 4ff_\mathit{eff}
        }
        {
        \bigg|
        \left(\frac{w}{2}\right)^2(-4f^2+x_m^2)+
        16f^2f_\mathit{eff}^2
        \bigg|
        }\\
        \approx&
        \hphantom{A}  
        \frac{x_m u_y w}{8f},
    \end{aligned}
    \end{equation}
    where we have made the same approximation as above.
    
    \item laterally scanning \underline{$y$} by setting \underline{$u_x=0$} and varying only \underline{$u_y$}, and analyzing the 1D geometric spot size in \underline{$x$} by propagating two marginal rays at entry positions \underline{$(x_m'=x_m+w/2,y_m'=0)$} and \underline{$(x_m'=x_m-w/2,y_m'=0)$}:
    \begin{equation}
    \begin{aligned}
        \delta_{g,x}(u_y,x_m)\propto
        &\hphantom{A}  
        \big|
        \widetilde{\mathbf{r}}_f^{(1)}\left(\mathbf{u}=(0,u_y),\mathbf{r}_m=\left(x_m,0\right),\mathbf{r}_m'=\left(x_m+w/2,0\right)\right)-\\
        &\hphantom{A}\hphantom{\big|}  
        \widetilde{\mathbf{r}}_f^{(1)}\left(\mathbf{u}=(0,u_y),\mathbf{r}_m=\left(x_m,0\right),\mathbf{r}_m'=\left(x_m-w/2,0\right)\right)
        \big|\\
        \propto&
        \frac{x_m u_y w}{8f}.
    \end{aligned}
    \end{equation}
    Here, the approximation was not necessary.
    
     \item laterally scanning \underline{$x$} by setting \underline{$u_y=0$} and varying only \underline{$u_x$}, and analyzing the 1D geometric spot size in \underline{$y$} by propagating two marginal rays at entry positions \underline{$(x_m'=x_m,y_m'=w/2)$} and \underline{$(x_m'=x_m,y_m'=-w/2)$}. 
    \begin{equation}\label{case4}
    \begin{aligned}
        \delta_{g,y}(u_x,x_m)\propto
        &\hphantom{A}  
        \big|
        \widetilde{\mathbf{r}}_f^{(1)}\left(\mathbf{u}=(u_x,0),\mathbf{r}_m=\left(x_m,0\right),\mathbf{r}_m'=\left(x_m,w/2\right)\right)-\\
        &\hphantom{A}\hphantom{\big|}  
        \widetilde{\mathbf{r}}_f^{(1)}\left(\mathbf{u}=(u_x,0),\mathbf{r}_m=\left(x_m,0\right),\mathbf{r}_m'=\left(x_m,-w/2\right)\right)
        \big|\\
        \propto&
        \hphantom{A}  
        x_mf_\mathit{eff}u_x\frac{w}{2}
        \frac{
        \left(\frac{w}{2}\right)^2 + 4ff_\mathit{eff}
        }
        {
        \bigg|
        \left(\frac{w}{2}\right)^2(-4f^2+x_m^2)+
        16f^2f_\mathit{eff}^2
        \bigg|
        }\\
        \approx&
        \hphantom{A}  
        \frac{x_m u_x w}{8f},
    \end{aligned}
    \end{equation}
    which we note is the same result as in case 2 above, with $x$ and $y$ swapped everywhere.
\end{enumerate}

Interestingly, all four cases give approximately the same result, indicating an isotropic spot size that linearly increases with lateral scan range. To get Eqs. \ref{case1}-\ref{case4} in terms of lateral position, we compute the distance of the central chief ray to the parabolic mirror focus (here, shifted to the origin):
\begin{equation}\label{lateral_pos}
    \big|
        \widetilde{\mathbf{r}}_f^{(1)}\left(\mathbf{u}=(u_x,u_y),\mathbf{r}_m=\left(x_m,0\right),\mathbf{r}_m'=\left(x_m,0\right)\right)
        \big|=f_\mathit{eff}\sqrt{u_x^2+u_y^2}.
\end{equation}
Thus, in this first-order approximation, the geometric spot size increases linearly with lateral position with slope
\begin{equation}
    \beta \propto \frac{x_m w}{ff_\mathit{eff}}.
\end{equation}
This is obtained by taking the ratios of Eqs. \ref{case1}-\ref{case4} and Eq. \ref{lateral_pos}, appropriately setting $u_x=0$ or $u_y=0$.

\subsection*{Geometric spot size when $x_m=0$}
We need a separate derivation for the second case, for which we start by performing a second-order Taylor expansion of Eq. \ref{final_ray} at $(u_x=0,u_y=0)$,
\begin{equation}\label{ray_2nd_order}
    \mathbf{r}_f(\mathbf{u},\mathbf{r}_m,\mathbf{r}_m')=
    \widetilde{\mathbf{r}}_f^{(2)}(\mathbf{u},\mathbf{r}_m,\mathbf{r}_m')
    +
    O\left((u_x+u_y)^3\right),
\end{equation}
where as before the superscript (2) indicates that this is a second-order approximation. As with the case of $x_m\gg0$ above, we propagate marginal rays to estimate the lateral spot size at the focal plane, except setting $x_m=0$. Due to symmetry, the four cases reduce to two cases here (i.e., cases 1 and 2 are the same, and cases 3 and 4 are the same):
\begin{enumerate}
    \item laterally scanning $x$ or $y$ and analyzing the 1D geometric spot size in the same direction:
    \begin{equation}\label{nonzero}
    \begin{aligned}
        \delta_{g,x}(u_x)\propto
        &\hphantom{A}  
        \big|
        \widetilde{\mathbf{r}}_f^{(2)}\left(\mathbf{u}=(u_x,0),\mathbf{r}_m=\left(0,0\right),\mathbf{r}_m'=\left(w/2,0\right)\right)-\\
        &\hphantom{A}\hphantom{\big|}  
        \widetilde{\mathbf{r}}_f^{(2)}\left(\mathbf{u}=(u_x,0),\mathbf{r}_m=\left(0,0\right),\mathbf{r}_m'=\left(-w/2,0\right)\right)
        \big|\\
        =&
        \hphantom{A}  
        w u_x^2\left(\frac{\frac{w^2}{4}+4f^2}{\frac{w^2}{4}-4f^2}\right)^2
        \approx w u_x^2,
    \end{aligned}
    \end{equation}
    again assuming $f\gg w$.
    \item laterally scanning $x$ or $y$ and analyzing the 1D geometric spot size in the orthogonal direction:
    \begin{equation}\label{zero}
    \begin{aligned}
        \delta_{g,x}(u_x)\propto
        &\hphantom{A}  
        \big|
        \widetilde{\mathbf{r}}_f^{(2)}\left(\mathbf{u}=(u_x,0),\mathbf{r}_m=\left(0,0\right),\mathbf{r}_m'=\left(w/2,0\right)\right)-\\
        &\hphantom{A}\hphantom{\big|}  
        \widetilde{\mathbf{r}}_f^{(2)}\left(\mathbf{u}=(u_x,0),\mathbf{r}_m=\left(0,0\right),\mathbf{r}_m'=\left(-w/2,0\right)\right)
        \big|\\
        =&
        \hphantom{A}  
        0.
    \end{aligned}
    \end{equation}
\end{enumerate}
Similarly to Eq. \ref{lateral_pos}, the lateral position is given by
\begin{equation}\label{lateral_pos2}
    \big|
        \widetilde{\mathbf{r}}_f^{(2)}\left(\mathbf{u}=(u_x,u_y),\mathbf{r}_m=\left(0,0\right),\mathbf{r}_m'=\left(0,0\right)\right)
        \big|=f\sqrt{u_x^2+u_y^2}.
\end{equation}
Thus, for this imaging geometry, using Eq. \ref{nonzero}, the geometric spot size increases radially quadratically with radial distance, with a coefficient of 
\begin{equation}
    \gamma \propto \frac{w}{f^2}.
\end{equation}
While Eq. \ref{zero} seems to indicate no geometric aberrations, this result is specific to the non-telecentric case considered in all of these derivations, where the angular scanning of the collimated beam pivots about the mirror surface. For the case of $x_m=0$, the degree of non-telecentricity affects the FOV, which is not covered by our derivations; however, for the case of $x_m\gg0$, telecentricity does not matter, as demonstrated through simulations in Sec. 5.7.

\section{Simulation results for ellipsoidal mirrors}
\begin{figure}
    \centering
    \includegraphics[width=\columnwidth]{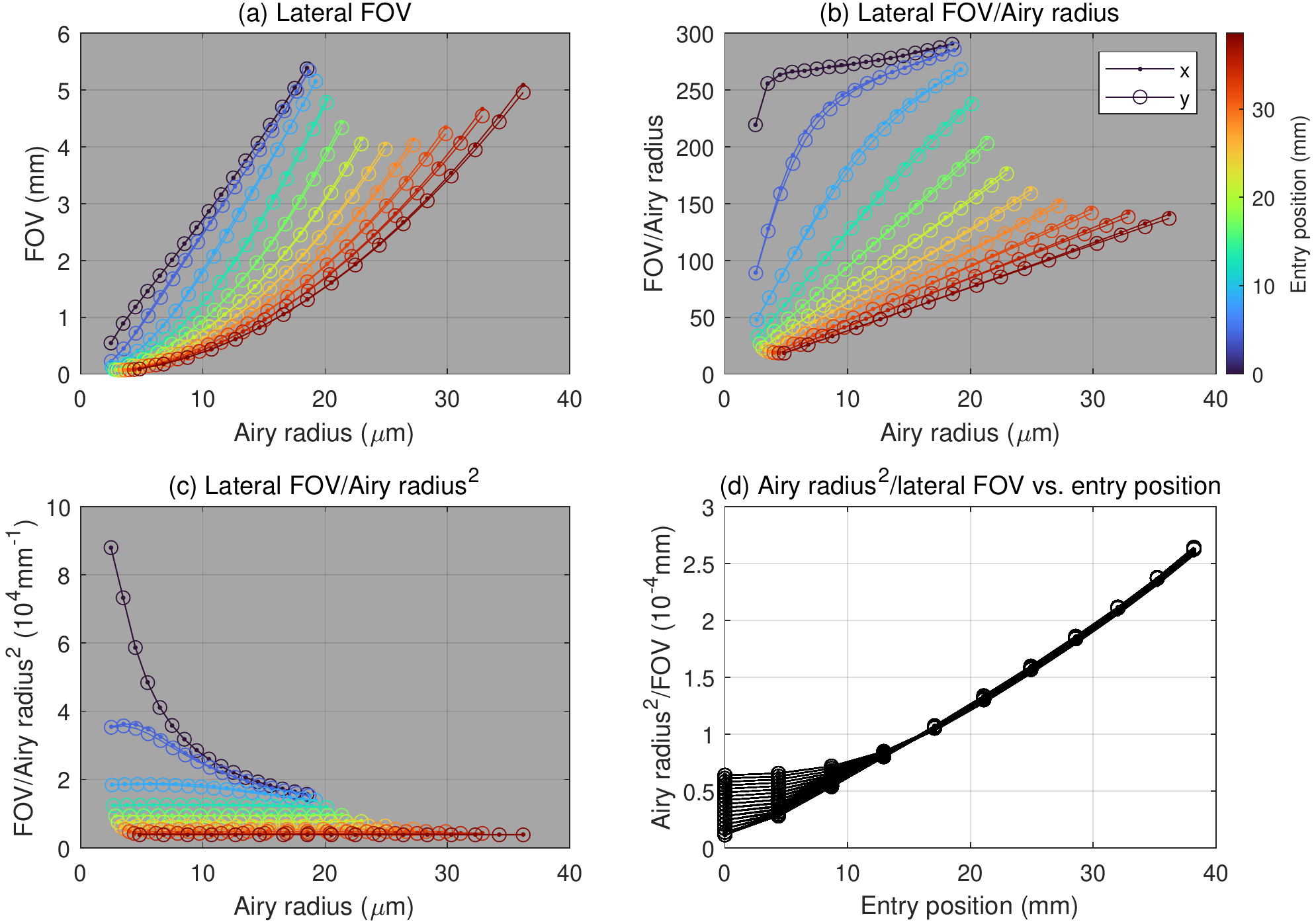}
    \caption{Lateral FOV scaling with lateral resolution at various entry positions ($x_m$) for an ellipsoidal mirror, simulated at $\lambda=800$ nm, $a=83.5$ mm, $b=56.48$ mm. We observe the same transition from quadratic to linear trends in (a)-(c) when $x_m\rightarrow0$ as for parabolic mirrors (Fig. 4). The relationship between $x_m$ and 1/FOV (d) is slightly super-linear.}
    \label{fig:FOV1_ellipsoidal}
\end{figure}

\begin{figure}
    \centering
    \includegraphics[width=.9\columnwidth]{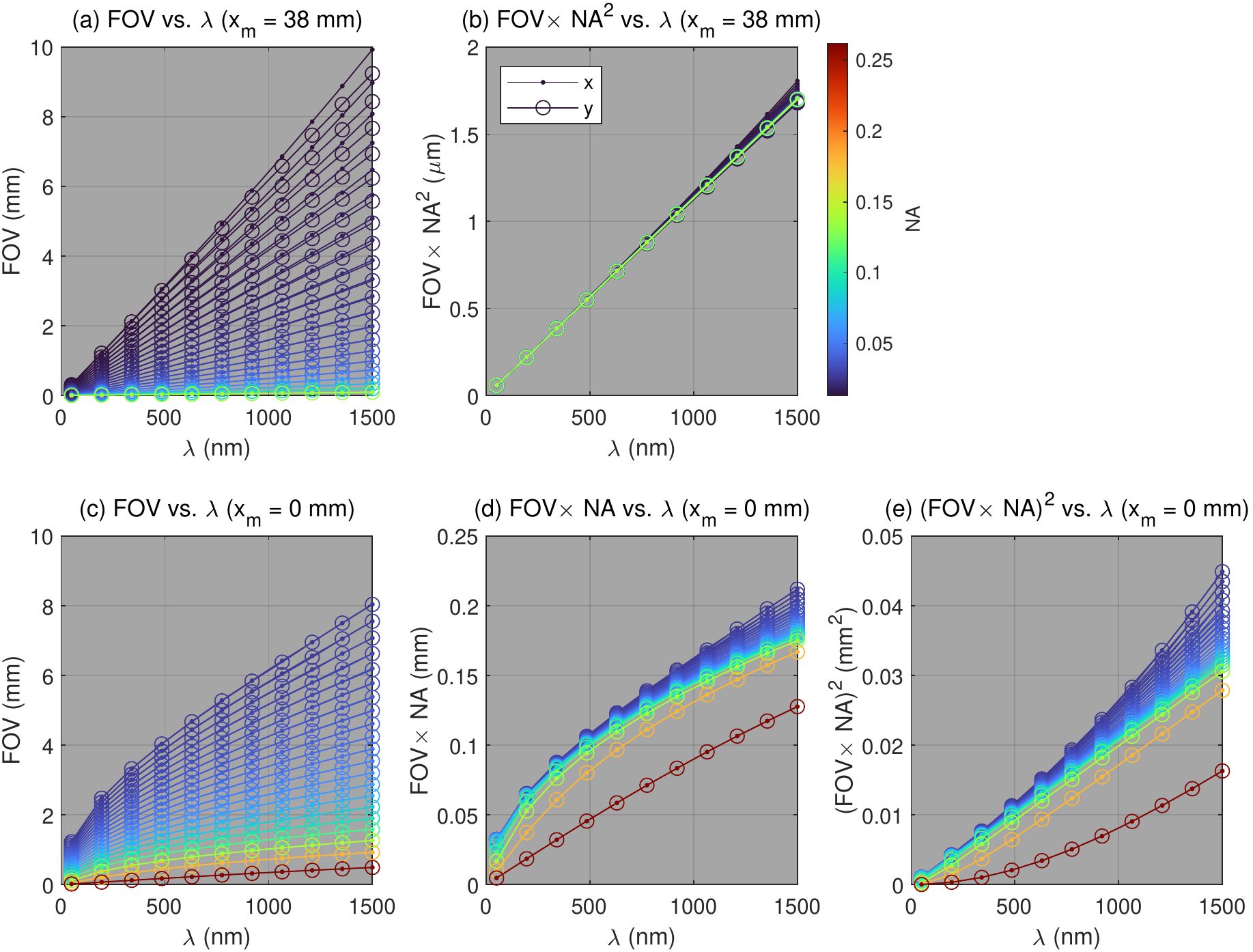}
    \caption{Lateral FOV scaling with wavelength ($\lambda$) for an ellipsoidal mirror, simulated at $a=83.5$ mm, $b=56.48$ mm and two entry positions, $x_m=38$ mm (first row) and $x_m=0$ mm (second row). The trends in all panels (a)-(e) are virtually identical to those for parabolic mirrors (see Fig. 5 for details).}
    \label{fig:FOV2_ellipsoidal}
\end{figure}

\begin{figure}
    \centering
    \includegraphics[width=\columnwidth]{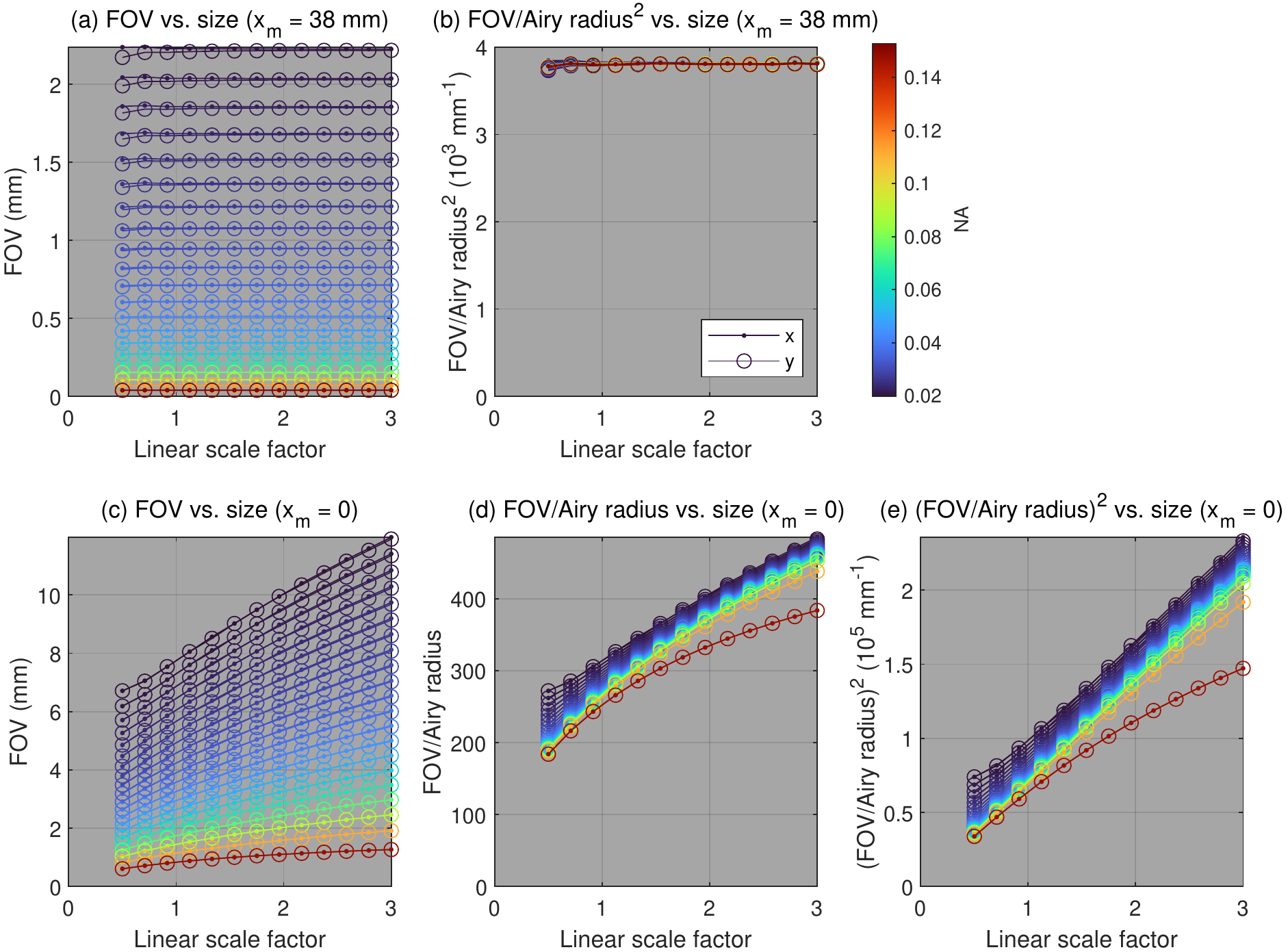}
    \caption{Lateral FOV scaling with ellipsoidal mirror size. Here, a linear scale factor of 1 corresponds to $a=83.5$ mm, $b=56.48$ mm. The trends here (a-e) are almost the same as for parabolic mirrors (Fig. 6).}
    \label{fig:FOV3_ellipsoidal}
\end{figure}

\begin{figure}
    \centering
    \includegraphics[width=\columnwidth]{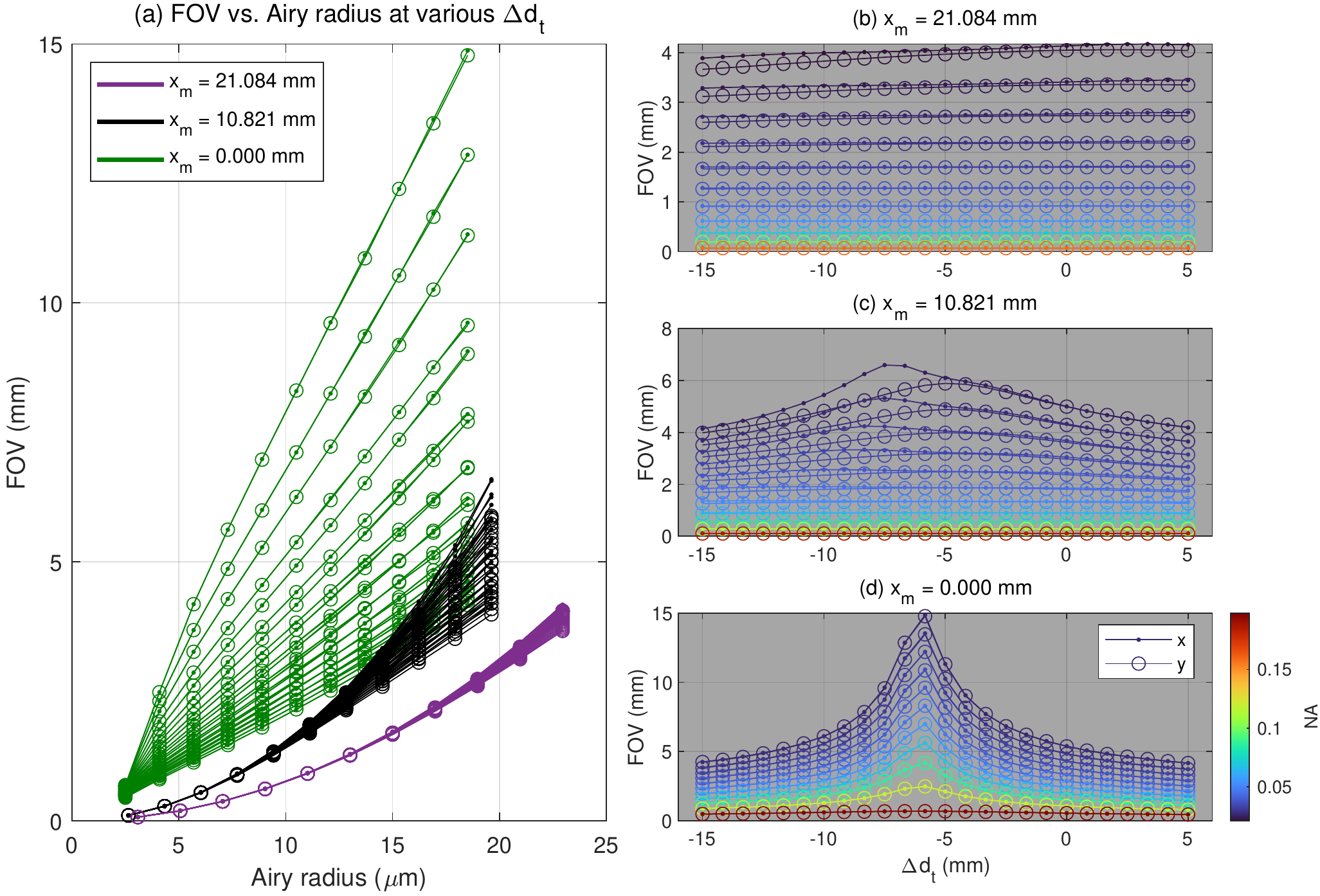}
    \caption{Effect telecentricity on lateral FOV for a ellipsoidal mirror, simulated at $\lambda=800$ nm and $a=83.5$ mm, $b=56.48$ mm. As in Fig. 7, $\Delta d_t=0$ corresponds to 90$^\circ$-sample-incidence configuration being telecentric. The trends here (a-d) are nearly identical to those of parabolic mirrors (Fig. 7). }
    \label{fig:FOV4_ellipsoidal}
\end{figure}

Figs. \ref{fig:FOV1_ellipsoidal}-\ref{fig:FOV4_ellipsoidal} show simulation results for ellipsoidal mirrors, analogous to those in Figs. 4-7 of the main text for parabolic mirrors. The trends are nearly identical for both types of mirrors.

\section{Zemax ray diagrams}
\begin{figure}
    \centering
    \includegraphics[width=\columnwidth]{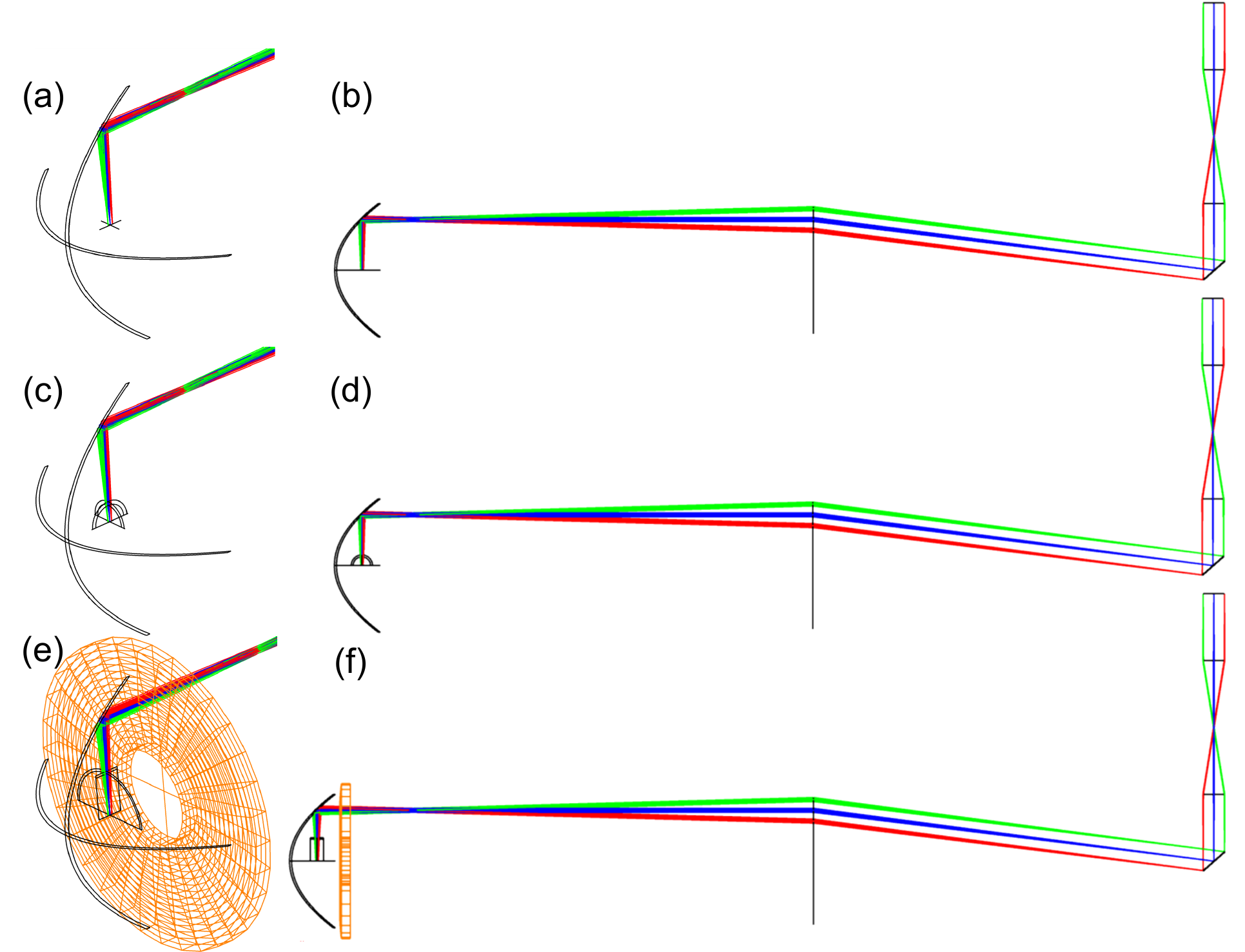}
    \caption{Zemax ray diagrams for three configurations of the parabolic mirror-based design: (a-b) sample in air, (c-d) sample in a water-filled optical dome, (e-f) sample in a water-filled cylindrical dish, with an astigmatism-correcting toroidal lens (orange) before the mirror. Panels (a), (c), and (e) show 3D views of the last few elements before the sample; panels (b), (d), and (f) show overviews of the entire design. All refractive elements before the parabolic mirror (except the toroidal lens) are ideal paraxial lenses.}
    \label{fig:parabolic_zemax}
\end{figure}

\begin{figure}
    \centering
    \includegraphics[width=.8\columnwidth]{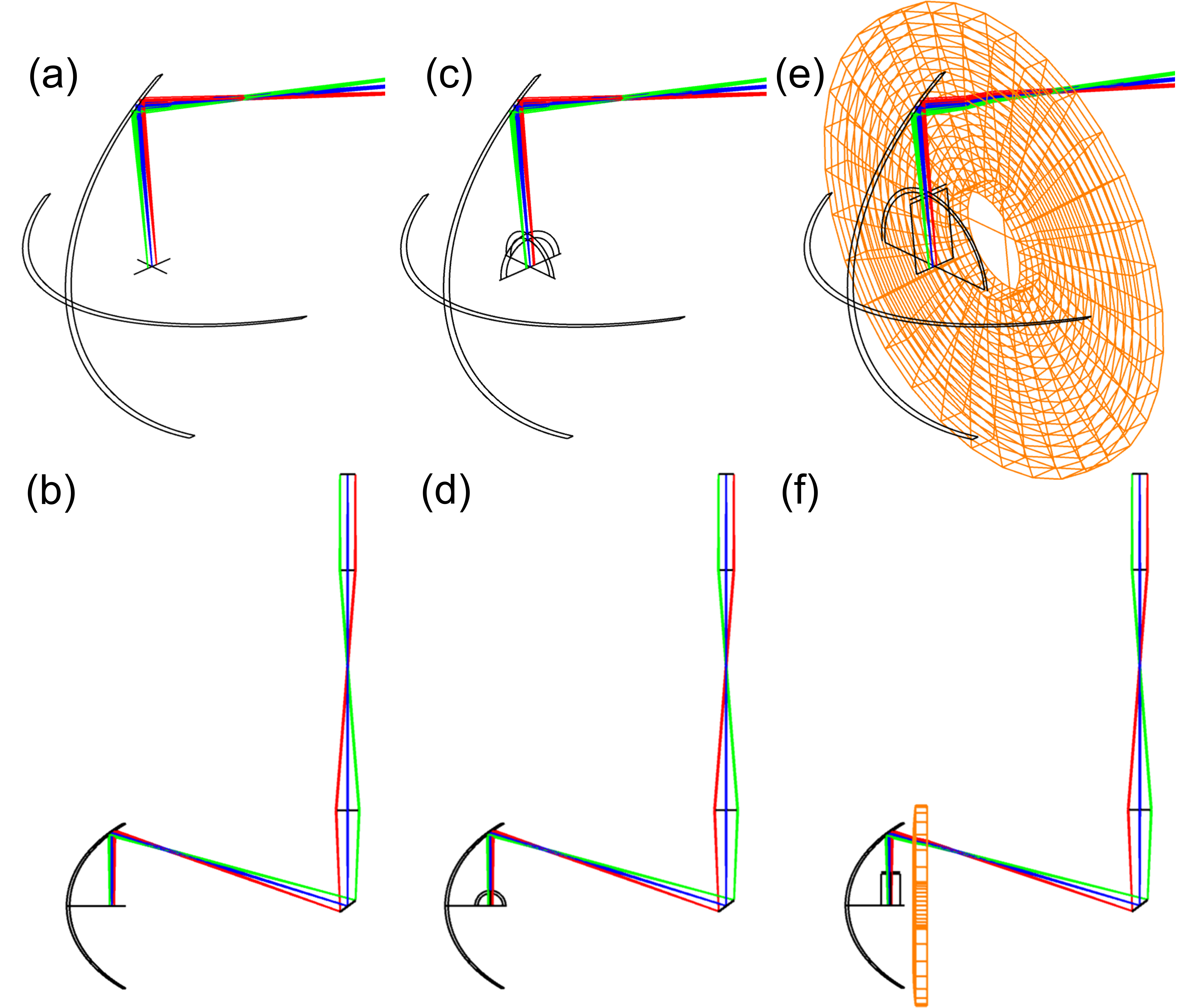}
    \caption{Zemax ray diagrams for three configurations of the ellipsoidal mirror-based design: (a-b) sample in air, (c-d) sample in a water-filled optical dome, (e-f) sample in a water-filled cylindrical dish, with an astigmatism-correcting toroidal lens (orange) before the mirror. Panels (a), (c), and (e) show 3D views of the last few elements before the sample; panels (b), (d), and (f) show overviews of the entire design. All refractive elements before the ellipsoidal mirror (except the toroidal lens) are ideal paraxial lenses. Note the lack of a tube lens compared to the parabolic design in Fig. \ref{fig:parabolic_zemax}.}
    \label{fig:ellipsoidal_zemax}
\end{figure}

Figs. \ref{fig:parabolic_zemax} and \ref{fig:ellipsoidal_zemax} show sample Zemax ray diagrams for the parabolic mirror-based and ellipsoidal mirror-based designs. See the accompanying Zemax files for detailed optical design parameters.

\clearpage

